\documentclass[english,eqsecnum,nofootinbib,superscriptaddress]{revtex4-2}
\usepackage[T1]{fontenc}
\usepackage[utf8]{inputenc}
\usepackage{color}
\usepackage{babel}
\usepackage{mathtools}
\usepackage{amsmath}
\usepackage[a4paper]{geometry}
\usepackage[pdfusetitle,
 bookmarks=true,bookmarksnumbered=false,bookmarksopen=false,
 breaklinks=false,pdfborder={0 0 1},backref=false,colorlinks=true]
 {hyperref}

\makeatletter
\usepackage{xcolor}
\usepackage{datetime2}
\usepackage{extarrows}
\makeatother

\begin{document}
\title{On the degrees of freedom of spatially covariant vector field theory}
\author{Shu-Yu Li}
\affiliation{School of Physics, Sun Yat-sen University, Guangzhou 510275, China}
\author{Xian Gao}
\email[Corresponding author: ]{gaoxian@mail.sysu.edu.cn}

\affiliation{School of Physics, Sun Yat-sen University, Guangzhou 510275, China}
\affiliation{Guangdong Provincial Key Laboratory of Quantum Metrology and Sensing,
Sun Yat-sen University, Zhuhai 519082, China}
\begin{abstract}
We investigate a class of spatially covariant vector field theories on a flat background, where the Lagrangians are constructed as polynomials of first-order derivatives of the vector field. Because Lorentz and $\mathrm{U}(1)$ invariances are broken, such theories generally propagate three degrees of freedom (DOFs): two transverse modes and one longitudinal mode. We examine the conditions under which the additional longitudinal mode is eliminated so that only two DOFs remain. To this end, we perform a Hamiltonian constraint analysis and identify two necessary and sufficient degeneracy conditions that reduce the number of DOFs from three to two. We find three classes of solutions satisfying these degeneracy conditions, corresponding to distinct types of theories. Type-I theories possess one first-class and two second-class constraints, type-II theories have four second-class constraints, and type-III theories contain two first-class constraints. The Maxwell theory is recovered as a special case of the type-III theories, where Lorentz symmetry is restored.
\end{abstract}
\maketitle

\section{Introduction}

Modifying general relativity (GR) to address unresolved problems such
as the cosmological constant and dark matter has become a common approach.
Lovelock's theorem shows that GR is the unique metric theory in four
dimensions whose equations of motion are second order, local, and
invariant under spacetime diffeomorphism \citep{Lovelock:1971yv}.
Consequently, any modification of GR must violate at least one of
these assumptions, for instance by introducing extra fields (as in
scalar-tensor theories), new geometric structures (such as non-Riemannian
geometry), higher dimensions, or higher-order derivatives. Among these
possibilities, scalar-tensor theories have seen remarkable progress
in the past two decades, especially in the construction of higher-derivative
models \citep{Horndeski:1974wa,Nicolis:2008in,Deffayet:2011gz,Kobayashi:2011nu,Langlois:2015cwa}.
However, single-field scalar-tensor theories with second-order derivatives
are now strongly constrained by observations, including those concerning
the propagation speed of gravitational waves \citep{LIGOScientific:2017vwq,Bettoni:2016mij,Ezquiaga:2017ekz,Sakstein:2017xjx,Baker:2017hug,Creminelli:2017sry}. 

A natural alternative is to introduce vector fields. Indeed, all fundamental
forces in the Standard Model are mediated by vector fields, which
motivates exploring their role in cosmology beyond scalar degrees
of freedom. Inspired by the success of constructing higher derivative
scalar-tensor theories, vector field theories with at most second-order
equations of motion have been developed \citep{Tasinato:2014eka}
and generalized to curved spacetimes under the name generalized Proca
theory \citep{Heisenberg:2014rta} (see also \citep{Allys:2015sht}).
The generalized Proca theory has been extensively studied in cosmology
\citep{DeFelice:2016yws,Heisenberg:2016eld,DeFelice:2016uil,BeltranJimenez:2016rff,DeFelice:2016cri},
in black hole physics \citep{Chagoya:2016aar,Babichev:2017rti,Heisenberg:2017hwb,Heisenberg:2017xda,Minamitsuji:2021gcq,Garcia-Saenz:2021uyv}
and in the context of gravitational waves \citep{Dong:2023xyb,Lai:2024fza,Dong:2024zal}.
See also \citep{Heisenberg:2018vsk} for a review and more references
therein. The generalized Proca theory has been further generalized
to the so-called extended vector-tensor theory, which relies on the
degeneracy of higher order derivatives rather than restricting the
order of the equations of motion \citep{Kimura:2016rzw,GallegoCadavid:2019zke,GallegoCadavid:2020dho,GallegoCadavid:2021ljh,deRham:2021efp,Aoki:2021wew,Aoki:2023bmz}.

A no-go theorem \citep{Deffayet:2013tca} forbids the Galileon-type
interactions of vector fields in flat spacetime if both Lorentz invariance
and $\mathrm{U}(1)$ gauge invariance are preserved. Indeed, in generalized
Proca theory and extended vector-tensor theory, $\mathrm{U}(1)$ invariance
is explicitly broken. Nevertheless, progress has been made in constructing
Einstein-Maxwell theories that retain $\mathrm{U}(1)$ symmetry \citep{Colleaux:2023cqu,Colleaux:2024ndy,Colleaux:2025vtm}.
Generally, once $\mathrm{U}(1)$ invariance is broken, the vector
sector propagates three DOFs: two transverse polarization modes and
an additional longitudinal mode.

The purpose of this work is to ask a basic question: is it possible
to construct a vector field theory that propagates only the two transverse
DOFs\footnote{This is also inspired by the recent progress of building gravitational
theories with only two DOFs such as \citep{Afshordi:2006ad,Lin:2017oow,Gao:2019twq}.}? Clearly, enforcing both Lorentz invariance and $\mathrm{U}(1)$
invariance recovers the Maxwell theory for a free vector field. Therefore,
to explore possible generalizations, we relax these symmetries and
require only spatial covariance, without assuming $\mathrm{U}(1)$
invariance. As a first step, we neglect gravitational effects and
work in flat spacetime. This setup can be viewed as a local limit
of a foliated cosmological spacetime. Locally the metric is approximately
flat on scales much smaller than the curvature scale, while a cosmological
evolution can still motivate a preferred time slicing and hence the
breaking of time diffeomorphisms. The relevant constraint structure
in the vector sector can still be captured in this setup. The analysis
in this work therefore serves as a ``decoupling limit'' of the theory
with gravitational effects. 

This flat-background analysis should be viewed as a controlled limit
of a preferred-foliation cosmological setting: locally (or in the
high-frequency/decoupling regime) the background metric can be approximated
by Minkowski while the reduced symmetry structure---spatial covariance
with broken time diffeomorphisms---still captures the relevant constraint
mechanism in the vector sector. 

In the context of the generalized Proca theory, the Faddeev-Jackiw
formalism has been applied to count the number of DOFs \citep{Sanongkhun:2019ntn}.
The so-called Maxwell-Proca theory proposed in \citep{ErrastiDiez:2019ttn,ErrastiDiez:2019trb}
shows that a vector field propagating two DOFs corresponds to a real
Abelian vector field whose Lagrangian involves two first-class constraints.
This naturally raises the question of whether one can construct a
more general flat-spacetime vector field theory that still propagates
two DOFs but involves additional second-class constraints. The present
work is devoted to exploring this question.

The remainder of the paper is organized as follows. In Sec. \ref{sec:vfsc},
we introduce our model of vector field theory respecting only the
spatial invariance. In Sec. \ref{sec:hampri} we set up our formalism
for the Hamiltonian constraint analysis and show that generally the
theory propagates three DOFs. In Sec. \ref{sec:dgncon1}, we derive
the first degeneracy condition in order to reduce the number of DOFs.
In Sec. \ref{sec:dgncon2} is devoted to the second degeneracy condition,
which is necessary in order to fully reduce the number of DOFs from
three to two. In Sec. \ref{sec:con}, we summarize our results.

\section{The spatially covariant vector field theory \label{sec:vfsc}}

Our purpose is to construct a vector field theory respecting only
spatial covariance. In this work, we neglect the gravitational effects
and consider a flat spacetime background, where the metric is 
\begin{equation}
\mathrm{d}s^{2}=-\mathrm{d}t^{2}+\delta_{ij}\mathrm{d}x^{i}\mathrm{d}x^{j}.
\end{equation}
That is, we treat the vector field $A_{\mu}$ as a test field in a
fixed flat background.

The global Lorentz symmetry is explicitly broken, and we retain only
the subgroup of spatial rotations $\mathrm{SO}(3)$. Accordingly,
we decompose the vector field into its temporal and spatial components,
\begin{equation}
A_{\mu}=\left(A_{0},A_{i}\right),
\end{equation}
where $i,j=1,2,3$ are spatial indices. Since Lorentz symmetry is
broken, the usual electromagnetic tensor $F_{\mu\nu}$ with components
\begin{equation}
F_{0i}=\dot{A}_{i}-\partial_{i}A_{0},\quad F_{ij}=\partial_{i}A_{j}-\partial_{j}A_{i},
\end{equation}
is no longer a preferred object. Instead, the Lagrangian can be a
general function of the first derivatives of the vector field. From
this perspective, the Maxwell theory corresponds to a particular combination
of quadratic monomials in the first derivatives of the vector field.

The most general spatially covariant Lagrangian takes the schematic
form $\mathcal{L}\left(A_{0},A_{i},\partial_{t},\partial_{i}\right).$
In this work, we restrict ourselves to Lagrangians that are polynomials
in the derivative terms. Each monomial is constructed from spatial
scalar formed by contracting the vector field and its first order
derivatives
\begin{equation}
\dot{A}_{0},\quad\partial_{i}A_{0},\quad\dot{A}_{i},\quad\partial_{i}A_{j},\label{1dvec}
\end{equation}
where an over-dot denotes a time derivative.

We will construct and classify possible monomials according to the
total number of derivatives. In this work, we focus on Lagrangians
up to the quadratic order in derivatives. That is, the total number
of derivatives in each monomial in the Lagrangian is at most two.

At the zeroth order in derivatives, the two basic spatial scalars
are $\bar{X}\equiv A_{i}A^{i}$ and $A_{0}$. The corresponding Lagrangian
is therefore an arbitrary function of these two quantities, i.e.,
\begin{equation}
\mathcal{L}_{2}=f_{2}(A_{0},\bar{X}).\label{Lag2}
\end{equation}
In this work we choose the subscripts of $\mathcal{L}_{n}$ following
the notation used in Horndeski and generalized Proca theories \citep{Deffayet:2011gz,Kobayashi:2011nu,Heisenberg:2014rta}\footnote{For example, in Horndeski theory, $\mathcal{L}_{2}$ denotes Lagrangian
involving no second derivative term of the scalar field, while $\mathcal{L}_{3}$
denotes Lagrangian that is linear in the second order derivative term
of the scalar field, etc.}.

At linear order in derivatives, there are five scalar contractions:
\[
\dot{A}_{0},\quad A^{i}\partial_{i}A_{0},\quad A^{i}\dot{A}_{i},\quad\partial_{i}A^{i},\quad A^{i}A^{j}\partial_{i}A_{j}.
\]
In this work, we require that contractions occur only among derivative
terms themselves. Terms involving contractions between derivatives
and the undifferentiated vector field, such as $A^{i}\partial_{i}A_{0}$,
$A^{i}\dot{A}_{i}$ and $A^{i}A^{j}\partial_{i}A_{j}$ will not be
considered. This restriction is primarily adopted for tractability
and for isolating the degeneracy structure in the simplest nontrivial
operator basis. Allowing additional operators (such as $A^{i}\partial_{i}A_{0}$,
$A^{i}\dot{A}_{i}$ or $A^{i}A^{j}\partial_{i}A_{j}$) enlarges the
basis substantially and in general modifies the momentum dependence
of the primary constraint and the subsequent constraint chain, so
that additional degeneracy conditions may be required to eliminate
the longitudinal mode. The Hamiltonian strategy developed in this
work can in principle be extended to the fully general basis, but
the systematic classification becomes considerably broader. The present
work therefore focuses on the above minimal class to obtain explicit,
classifiable two-DOF theories. With this assumption, we are left with
two admissible monomials, $\dot{A}_{0}$ and $\partial_{i}A^{i}$,
and the Lagrangian is given by
\begin{equation}
\mathcal{L}_{3}=f_{3,1}\dot{A}_{0}+f_{3,2}\partial_{i}A^{i},\label{Lag3}
\end{equation}
where coefficients $f_{3,1}$ and $f_{3,2}$ are general functions
of $A_{0}$ and $\bar{X}$.

As quadratic order in derivatives, one can form a large number of
scalar monomials by contracting indices of $\dot{A}_{0}$, $\partial_{i}A_{0}$,
$\dot{A}_{i}$ and $\partial_{i}A_{j}$. We find ten types of monomial.
The complete set of scalar monomials can be classified as:
\begin{itemize}
\item $\left(\dot{A}_{0}\right)^{2}$
\item $A^{i}\dot{A}_{0}\partial_{i}A_{0}$
\item $\dot{A}_{0}\dot{A}_{i}A^{i}$
\item $\dot{A}_{0}\partial_{i}A^{i}$, $\dot{A}_{0}\partial_{i}A_{j}A^{i}A^{j}$
\item $\partial^{i}A_{0}\partial_{i}A_{0}$, $A^{i}A^{j}\partial_{j}A_{0}\partial_{i}A_{0}$
\item $\partial^{i}A_{0}\dot{A}_{i}$, $A^{i}A^{j}\partial_{j}A_{0}\dot{A}_{i}$
\item $A^{k}\partial_{k}A_{0}\partial_{i}A^{i}$, $A^{i}\partial^{j}A_{0}\partial_{i}A_{j}$,
$A^{j}\partial^{i}A_{0}\partial_{i}A_{j}$, $A^{k}A^{i}A^{j}\partial_{k}A_{0}\partial_{i}A_{j}$
\item $\dot{A}_{i}\dot{A}^{i}$, $A^{i}A^{j}\dot{A}_{j}\dot{A}_{i}$
\item $A^{k}\dot{A}_{k}\partial_{i}A^{i}$, $A^{i}\dot{A}^{j}\partial_{i}A_{j}$,
$A^{j}\dot{A}^{i}\partial_{i}A_{j}$, $A^{k}A^{i}A^{j}\dot{A}_{k}\partial_{i}A_{j}$
\item $\partial_{i}A^{i}\partial_{j}A^{j}$, $\partial_{i}A_{j}\partial^{i}A^{j}$,
$\partial_{i}A_{j}\partial^{j}A^{i}$, $A^{i}A^{j}\partial_{k}A^{k}\partial_{i}A_{j}$,
$A^{i}A^{j}\partial_{k}A_{i}\partial^{k}A_{j}$, \\$A^{i}A^{j}\partial_{k}A_{i}\partial_{j}A^{k}$,
$A^{i}A^{j}\partial_{i}A_{k}\partial_{j}A^{k}$, $A^{i}A^{j}A^{k}A^{l}\partial_{i}A_{j}\partial_{k}A_{l}$
\end{itemize}
As in the case of the linear order, if we impose the same restriction
that contractions must occur only among derivative terms, we are left
with the following eight contractions
\begin{align*}
 & \left(\dot{A}_{0}\right)^{2},\quad\dot{A}_{0}\partial_{i}A^{i},\quad\partial^{i}A_{0}\partial_{i}A_{0},\quad\partial^{i}A_{0}\dot{A}_{i},\\
 & \dot{A}_{i}\dot{A}^{i},\quad\partial_{i}A^{i}\partial_{j}A^{j},\quad\partial_{i}A_{j}\partial^{i}A^{j},\quad\partial_{i}A_{j}\partial^{j}A^{i}.
\end{align*}
As a result, the Lagrangian at the quadratic order in derivatives
is given by
\begin{align}
\mathcal{L}_{4}= & f_{4,1}(\dot{A}_{0})^{2}+f_{4,2}\dot{A}_{0}\partial_{i}A^{i}\nonumber \\
 & +h_{4,1}\dot{A}_{i}\dot{A}^{i}+h_{4,2}\partial_{i}A_{0}\partial^{i}A_{0}+h_{4,3}\dot{A}_{i}\partial^{i}A_{0}\nonumber \\
 & \begin{aligned} & +f_{4,3}(\partial_{i}A^{i})^{2}+f_{4,4}\partial_{i}A_{j}\partial^{i}A^{j}+f_{4,5}\partial_{i}A_{j}\partial^{j}A^{i},\end{aligned}
\label{Lag4}
\end{align}
where again all coefficients are general functions of $A_{0}$ and
$\bar{X}$, i.e., $f_{m,n}=f_{m,n}(A_{0},\bar{X})$ and $h_{m,n}=h_{m,n}(A_{0},\bar{X})$.
In the second line of (\ref{Lag4}), we denote the corresponding coefficients
by $h_{m,n}$ purely for notational convenience. No physical distinction
is implied: all $f_{m,n}$ and $h_{m,n}$ are general functions of
$A_{0}$ and $\bar{X}$. The split only helps us track the operators
involving the $\dot{A}_{i}$-type and $\partial_{i}A_{0}$-type contractions
when deriving conjugate momenta and degeneracy conditions.

Generally, this procedure can be extended to construct Lagrangians
at higher orders in derivatives. In this work we focus on the Lagrangian
up to the quadratic order in derivative terms. Finally, the model
we are considering in this work is 
\begin{equation}
S=\int\mathrm{d}t\mathrm{d}^{3}x\,\mathcal{L}=\int\mathrm{d}t\mathrm{d}^{3}x\left(\mathcal{L}_{2}+\mathcal{L}_{3}+\mathcal{L}_{4}\right),\label{action}
\end{equation}
where $\mathcal{L}_{2}$, $\mathcal{L}_{3}$ and $\mathcal{L}_{4}$
are given in (\ref{Lag2}), (\ref{Lag3}) and (\ref{Lag4}), respectively.

\section{Hamiltonian and primary constraints \label{sec:hampri}}

Our aim in this section is to determine the number of degrees of freedom
(DOFs) carried by the action in (\ref{action}). We begin by examining
the Hessian matrix of the kinetic terms $\dot{A}_{\mu}=\{\dot{A}_{0},\dot{A}_{i}\}$,
which is defined by
\begin{equation}
\mathcal{H}^{\mu\nu}\coloneqq\frac{\delta^{2}S}{\delta\dot{A}_{\mu}\delta\dot{A}_{\nu}}=\begin{pmatrix}2f_{4,1}\\
 & 2h_{4,1}\\
 &  & 2h_{4,1}\\
 &  &  & 2h_{4,1}
\end{pmatrix}.
\end{equation}
Clearly, if $\det\mathcal{H}^{\mu\nu}\neq0$, then all four components
of $A_{\mu}$ are dynamical, implying four propagating DOFs.

To find the condition under which the theory propagates only two DOFs,
a necessary condition is that the Hessian matrix is degenerate. In
order to have $\det\mathcal{H}^{\mu\nu}=0$, we have two possibilities:
\begin{enumerate}
\item $h_{4,1}=0$: In this case, three eigenvalues of the Hessian matrix
vanish, yielding three primary constraints. However, all spatial components
$A_{i}$ become non-dynamical with this choice, which is unphysical
as there will be no transverse polarization DOFs at all. We thus do
not consider this case.
\item $f_{4,1}=0$: In this case, the Hessian matrix possesses one vanishing
eigenvalue, corresponding to a single primary constraint. The temporal
component $A_{0}$ acts as an auxiliary variable, as in the standard
Maxwell theory.
\end{enumerate}
Henceforth we adopt the second choice, $f_{4,1}=0$. We will examine
the number of DOFs with this choice. 

Counting number of DOFs can be well performed in the Hamiltonian analysis.
With $f_{4,1}=0$, the canonical momenta defined by $\varPi^{\mu}=\delta\mathcal{L}/\delta\dot{A}_{\mu}$
are
\begin{align}
\varPi^{0} & =f_{3,1}+f_{4,2}\partial_{i}A^{i},\label{Pi0}\\
\varPi^{i} & =2h_{4,1}\dot{A}^{i}+h_{4,3}\partial^{i}A_{0}.\label{Pii}
\end{align}
As expected, $\varPi^{0}$ in (\ref{Pi0}) contains no time derivatives
of the fields, leading to a primary constraint
\begin{equation}
\phi_{1}\equiv\varPi^{0}-(f_{4,2}\partial_{i}A^{i}+f_{3,1})\approx0.\label{phi1}
\end{equation}
Here and throughout this work, ``$\approx$'' represents ``weak
equality'', which holds on the constraint surface $\varGamma_{\mathrm{p}}$
defined by the constraints in phase space. 

The canonical Hamiltonian $\mathcal{H}_{\mathrm{C}}$ is defined as
usual by
\begin{equation}
\mathcal{H}_{\mathrm{C}}=\varPi^{0}\dot{A}_{0}+\varPi^{i}\dot{A}_{i}-\mathcal{L}.
\end{equation}
By solving $\dot{A}_{i}$ from (\ref{Pii}) and using the primary
constraint (\ref{phi1}), the canonical Hamiltonian can be expressed
as a function of $\varPi^{i}$, $A_{0}$ and $A_{i}$,
\begin{align}
\mathcal{H}_{\mathrm{C}} & =\mathcal{K}(\varPi^{i})^{2}+\mathcal{W}_{i}^{(0)}\varPi^{i}+\mathcal{V}^{(0,i)},\label{Ham_canon}
\end{align}
where $\mathcal{K}$, $\mathcal{W}_{i}^{(0)}$ and $\mathcal{V}^{(0,i)}$
are functions that do not contain $\varPi^{i}$. Here the superscripts
indicate which configuration variables the coefficients depend on.
That is, $(0)$ denotes dependence only on $A_{0}$ (and its spatial
derivatives), while $(0,i)$ denotes dependence on both $A_{0}$ and
$A_{i}$ (and their spatial derivatives). In (\ref{Ham_canon}), $\mathcal{K}$,
$\mathcal{W}_{i}^{(0)}$ and $\mathcal{V}^{(0,i)}$ are given by
\begin{equation}
\mathcal{K}=\frac{1}{4h_{4,1}},\label{calK}
\end{equation}
\begin{equation}
\mathcal{W}_{i}^{(0)}=-\frac{h_{4,3}}{2h_{4,1}}\partial_{i}A_{0},\label{calW0}
\end{equation}
and
\begin{align}
\mathcal{V}^{(0,i)}= & -f_{2}-f_{3,2}\partial_{i}A^{i}-\left(h_{4,2}-\frac{(h_{4,3})^{2}}{4h_{4,1}}\right)\partial_{i}A_{0}\partial^{i}A_{0}\nonumber \\
 & -f_{4,3}(\partial_{i}A^{i})^{2}-f_{4,4}\partial_{i}A_{j}\partial^{i}A^{j}-f_{4,5}\partial_{i}A_{j}\partial^{j}A^{i}.\label{calV0i}
\end{align}

In the presence of primary constraints, the time evolution is determined
by the total Hamiltonian defined by
\begin{equation}
H_{\mathrm{T}}:=H_{\mathrm{C}}+\int\mathrm{d}^{3}y\lambda^{I}(\vec{y})\phi_{I}(\vec{y}),
\end{equation}
where $\lambda^{I}$ are the undetermined Lagrange multipliers enforcing
the constraints $\phi_{I}$. Here and in what follows, summation over
indices $I,J$ is implicitly understood. Moreover, since the canonical
Hamiltonian $H_{\mathrm{C}}$ is well-defined only on the submanifold
$\varGamma_{\mathrm{p}}$ defined by the primary constraints, and
can be extended off $\varGamma_{\mathrm{p}}$, one can usually use
$H_{\mathrm{C}}|_{\varGamma_{\mathrm{p}}}$ instead of $H_{\mathrm{C}}$
when performing the explicit calculations.

The time evolution of any function $F$ of phase space variables (that
does not explicitly depend on $t$) is given by the Poisson bracket
\begin{equation}
\dot{F}=[F,H_{\mathrm{C}}]+\int\mathrm{d}^{3}y\,\lambda^{I}(\vec{y})[F,\phi_{I}(\vec{y})]\label{dotF}
\end{equation}
with $H_{\mathrm{C}}=\int\mathrm{d}^{3}x\mathcal{H}_{\mathrm{C}}(\vec{x})$,
where the Poisson bracket of any two functions $F,G$ is defined by
\begin{equation}
[F,G]=\int\mathrm{d}^{3}z\left(\frac{\delta F}{\delta A^{I}(\vec{z})}\frac{\delta G}{\delta\varPi_{I}(\vec{z})}-\frac{\delta F}{\delta\varPi_{I}(\vec{z})}\frac{\delta G}{\delta A^{I}(\vec{z})}\right).\label{PBdef}
\end{equation}
A basic consistency requirement is that the primary constraints be
preserved in time, i.e. $\dot{\phi}_{I}\approx0$. Applying (\ref{dotF})
yields the consistency conditions, 
\begin{equation}
\int\mathrm{d}^{3}y[\phi_{I}(\vec{x}),\phi_{J}(\vec{y})]\lambda^{J}(\vec{y})+[\phi_{I}(\vec{x}),H_{\mathrm{C}}]\approx0.\label{cons}
\end{equation}
Note that (\ref{cons}) may either determine the Lagrange multipliers
$\{\lambda^{I}\}$ or reduce to relations among the phase space variables
$\{A_{\mu},\varPi^{\mu}\}$. In the latter case, there may be secondary
constraints if the relations are independent of the primary constraints.
Hence, it is necessary to evaluate the Poisson brackets among the
primary constraints as well as the Poisson brackets between the primary
constraints and the canonical Hamiltonian.

In the present case, there is only one primary constraint $\phi_{1}(\vec{x})$,
the Poisson bracket between $\phi_{1}(\vec{x})$ and $\phi_{1}(\vec{y})$
is given by 
\begin{equation}
[\phi_{1}(\vec{x}),\phi_{1}(\vec{y})]=\int\mathrm{d}^{3}z\,\delta^{(3)}(y-z)\delta^{(3)}(x-z)\left[\frac{\partial f_{4,2}}{\partial A_{0}}\partial_{i}A^{i}(\vec{x})+\frac{\partial f_{3,1}}{\partial A_{0}}\right]-(x\leftrightarrow y).
\end{equation}
Since the delta function is symmetric under $x\leftrightarrow y$,
the Poisson bracket vanishes identically, i.e.,
\begin{equation}
[\phi_{1}(\vec{x}),\phi_{1}(\vec{y})]=0.\label{PBphi1}
\end{equation}

We must now verify whether $\dot{\phi}_{1}\approx0$ is automatically
satisfied or induces a secondary constraint. According to (\ref{dotF}),
we have
\[
\dot{\phi}_{1}(\vec{x})=[\phi_{1}(\vec{x}),H_{\mathrm{C}}]+\int\mathrm{d}^{3}y\,\lambda(\vec{y})[\phi_{1}(\vec{x}),\phi_{1}(\vec{y})]\equiv[\phi_{1}(\vec{x}),H_{\mathrm{C}}],
\]
where the second term vanishes due to (\ref{PBphi1}). Since $[\phi_{1}(\vec{x}),H_{\mathrm{C}}]$
cannot be expressed as a function of $\phi_{1}(\vec{x})$, the consistency
condition of $\phi_{1}$ yields a secondary constraint
\begin{equation}
\dot{\phi}_{1}=[\phi_{1}(\vec{x}),H_{\mathrm{C}}]=\mathcal{C}(\vec{x})\approx0.
\end{equation}
After some manipulations, the explicit expression for the secondary
constraint is found to be
\begin{equation}
\mathcal{C}\equiv\dot{\phi}_{1}=\mathcal{A}\partial_{i}\varPi^{i}+\mathcal{B}(\varPi^{i})^{2}+\mathcal{C}_{i}^{(0,i)}\varPi^{i}+\mathcal{D}^{(0,i)},
\end{equation}
where $(\varPi^{i})^{2}\equiv\varPi_{i}\varPi^{i}$ and $\mathcal{A},\mathcal{B},\mathcal{C}_{i},\mathcal{D}$
are functions involving no $\varPi^{i}$, given by
\begin{equation}
\mathcal{A}=-2\mathcal{K}(f_{4,2}+h_{4,3}),\label{calA}
\end{equation}
\begin{equation}
\mathcal{B}=-\frac{\partial\mathcal{K}}{\partial A_{0}},
\end{equation}
\begin{align}
\mathcal{C}_{i}^{(0,i)}= & \frac{\partial\mathcal{W}_{i}^{(0)}}{\partial A_{0}}-\frac{\partial}{\partial A_{0}}\left(\frac{h_{4,3}}{2h_{4,1}}\right)\partial_{i}A_{0}-\frac{\partial}{\partial A_{k}}\left(\frac{h_{4,3}}{2h_{4,1}}\right)\partial_{i}A_{k}\nonumber \\
 & -\left(\frac{\partial f_{4,2}}{\partial A_{k}}\partial_{l}A^{l}-\frac{\partial f_{4,2}}{\partial A_{0}}\partial^{k}A_{0}-\frac{\partial f_{4,2}}{\partial A_{l}}\partial^{k}A_{l}+\frac{\partial f_{3,1}}{\partial A_{k}}\right)2\mathcal{K}\delta_{ki}\nonumber \\
 & +2f_{4,2}\delta_{ki}\left(\frac{\partial\mathcal{K}}{\partial A_{0}}\partial^{k}A_{0}+\frac{\partial\mathcal{K}}{\partial A_{l}}\partial^{k}A_{l}\right),\label{calC0i}
\end{align}
and
\begin{align}
\mathcal{D}^{(0,i)}= & \frac{\partial\mathcal{V}^{(0,i)}}{\partial A_{0}}-2\frac{\partial}{\partial A_{0}}\left(h_{4,2}-\frac{(h_{4,3})^{2}}{4h_{4,1}}\right)\partial^{i}A_{0}\partial_{i}A_{0}-2\frac{\partial}{\partial A_{k}}\left(h_{4,2}-\frac{(h_{4,3})^{2}}{4h_{4,1}}\right)\partial^{i}A_{0}\partial_{i}A_{k}\nonumber \\
 & -2\left(h_{4,2}-\frac{(h_{4,3})^{2}}{4h_{4,1}}\right)\partial_{i}\partial^{i}A_{0}-\left(\frac{\partial f_{4,2}}{\partial A_{k}}\partial_{i}A^{i}-\frac{\partial f_{4,2}}{\partial A_{0}}\partial^{k}A_{0}-\frac{\partial f_{4,2}}{\partial A_{l}}\partial^{k}A_{l}+\frac{\partial f_{3,1}}{\partial A_{k}}\right)\mathcal{W}_{k}^{(0)}\nonumber \\
 & +f_{4,2}\left(\frac{\partial\mathcal{W}_{k}^{(0)}}{\partial A_{0}}\partial^{k}A_{0}+\frac{\partial\mathcal{W}_{k}^{(0)}}{\partial A_{l}}\partial^{k}A_{l}\right),\label{calD0i}
\end{align}
where $\mathcal{K}$, $\mathcal{W}_{i}^{(0)}$ and $\mathcal{V}^{(0,i)}$
are given by (\ref{calK})-(\ref{calV0i}).

We have now two constraints, the primary constraint $\phi_{1}$ and
its associated secondary constraint $\mathcal{C}\equiv\dot{\phi}_{1}$.
Without additional requirement, the Poisson bracket between $\phi_{1}$
and $\dot{\phi}_{1}$ does not vanish generally, and thus there will
be no further constraint\footnote{This can be seen explicitly from $\ddot{\phi}_{1}(\vec{x})=[\dot{\phi}_{1}(\vec{x}),H_{\mathrm{C}}]+\int\mathrm{d}^{3}y\,\lambda(\vec{y})[\dot{\phi}_{1}(\vec{x}),\phi_{1}(\vec{y})]$,
since $[\dot{\phi}_{1}(\vec{x}),\phi_{1}(\vec{y})]\neq0$ weakly,
the consistency condition of $\dot{\phi}_{1}$ merely fixes the Lagrange
multiplier $\lambda$ instead of yielding a tertiary constraint.}. The Dirac matrix formed by these constraints reads
\begin{equation}
A^{ab}(\vec{x},\vec{y})=\begin{array}{|c|cc|}
\hline [\cdot,\cdot] & \phi_{1} & \dot{\phi}_{1}\\
\hline \phi_{1} & 0 & X\\
\dot{\phi}_{1} & X & X
\\\hline \end{array},\label{Aab_dof3}
\end{equation}
where ``$X$'' stands for generally non-vanishing entries. According
to Dirac-Bergmann's terminology, both constraints are second class.
As a result, the number of DOFs is given by
\begin{equation}
\#_{\mathrm{DOF}}=\frac{2\times4_{\mathrm{var}}-1\times2_{2\mathrm{nd}}}{2}=3,
\end{equation}
as expected. 

In the following, we will examine under which condition the number
of DOFs can be less than 3, and in particular there are only 2 dynamical
DOFs propagating. For the sake of simplicity, we focus on the case
in which the coefficients in the Lagrangian (i.e., $f_{3,1}$, $f_{4,1}$
etc.) are polynomials of $A_{0}$ and $\bar{X}$. 

\section{The first degeneracy condition \label{sec:dgncon1}}

As discussed in the previous section, in order to have number of DOFs
less than 3, a necessary condition is that the Poisson bracket between
$\phi_{1}$ and $\dot{\phi}_{1}$ must be vanishing (at least weakly).
That is, we must require
\begin{equation}
[\phi_{1}(\vec{x}),\dot{\phi}_{1}(\vec{y})]\approx0,\label{firstdgc}
\end{equation}
which we refer to as the first degeneracy condition.

After a lengthy but straightforward computation, the Poisson bracket
can be written as
\begin{align}
[\phi_{1}(\vec{x}),\dot{\phi}_{1}(\vec{y})] & =\mathcal{F}_{1}(\vec{x})\delta^{(3)}(\vec{y}-\vec{x})+\mathcal{F}^{i}(\vec{x})\frac{\partial}{\partial y^{i}}\delta^{(3)}(\vec{y}-\vec{x})+\mathcal{F}_{2}(\vec{x})\frac{\partial^{2}}{\partial y^{i}\partial y_{i}}\delta^{(3)}(\vec{y}-\vec{x}),\label{PBphiphidot}
\end{align}
where the coefficients are given by
\begin{equation}
\mathcal{F}_{1}=-\frac{\partial\mathcal{A}}{\partial A_{0}}\partial_{i}\varPi^{i}-\frac{\partial\mathcal{B}}{\partial A_{0}}(\varPi^{i})^{2}-\frac{\partial\mathcal{C}_{i}^{(0,i)}}{\partial A_{0}}\varPi^{i}-\frac{\partial\mathcal{D}^{(0,i)}}{\partial A_{0}}+\left(\frac{\partial f_{4,2}}{\partial A_{k}}\partial_{i}A^{i}+\frac{\partial f_{3,1}}{\partial A_{k}}\right)(2\mathcal{B}\varPi_{k}+\mathcal{C}_{k}^{(0,i)}),
\end{equation}
\begin{align}
\mathcal{F}^{i}= & \left[\frac{\partial}{\partial A_{0}}\left(\frac{h_{4,3}}{2h_{4,1}}\right)+\frac{\partial}{\partial A_{0}}\left(\frac{h_{4,3}}{2h_{4,1}}\right)-2\frac{\partial f_{4,2}}{\partial A_{0}}\mathcal{K}+2f_{4,2}\frac{\partial\mathcal{K}}{\partial A_{0}}\right]\varPi^{i}+6\frac{\partial}{\partial A_{0}}\left(h_{4,2}-\frac{(h_{4,3})^{2}}{4h_{4,1}}\right)\partial^{i}A_{0}\nonumber \\
 & +\frac{\partial f_{4,2}}{\partial A_{0}}\mathcal{W}^{i(0)}+f_{4,2}\frac{\partial\mathcal{W}^{i(0)}}{\partial A_{0}}+\left(\frac{\partial f_{4,2}}{\partial A_{i}}\partial_{l}A^{l}+\frac{\partial f_{3,1}}{\partial A_{i}}\right)\mathcal{A}+(2\mathcal{B}\varPi^{i}+\mathcal{C}^{i(0,i)})f_{4,2}\nonumber \\
 & +\left[\frac{\partial}{\partial A_{0}}(f_{4,2}\mathcal{A})\partial^{i}A_{0}+\frac{\partial}{\partial A_{k}}(f_{4,2}\mathcal{A})\partial^{i}A_{k}\right],
\end{align}
and
\begin{equation}
\mathcal{F}_{2}=f_{4,2}\mathcal{A}+2h_{4,2}-\frac{(h_{4,3})^{2}}{2h_{4,1}}.
\end{equation}
In deriving these expressions, we have used integrations by parts.
For example, we have
\begin{align}
 & \int\mathrm{d}^{3}z\,f_{4,2}\mathcal{A}\partial_{x^{i}}\delta^{(3)}(\vec{x}-\vec{z})\partial^{y^{i}}\delta^{(3)}(\vec{y}-\vec{z})\nonumber \\
= & -f_{4,2}\mathcal{A}\partial_{x^{i}}\partial^{x^{i}}\delta^{(3)}(\vec{y}-\vec{x})-\left[\frac{\partial}{\partial A_{0}}(f_{4,2}\mathcal{A})\partial_{x^{i}}A_{0}+\frac{\partial}{\partial A_{k}}(f_{4,2}\mathcal{A})\partial_{x^{i}}A_{k}\right]\partial^{x^{i}}\delta^{(3)}(\vec{y}-\vec{x}).
\end{align}

In order to have a vanishing Poisson bracket (\ref{firstdgc}), we
must require that 
\begin{equation}
\mathcal{F}_{1}(\vec{x})\delta^{(3)}(\vec{y}-\vec{x})+\mathcal{F}^{i}(\vec{x})\frac{\partial\delta^{(3)}(\vec{y}-\vec{x})}{\partial y^{i}}+\mathcal{F}_{2}(\vec{x})\frac{\partial^{2}\delta^{(3)}(\vec{y}-\vec{x})}{\partial y^{i}\partial y_{i}}=0.\label{firstdgc_xpl}
\end{equation}
Note we have used ``strong'' equality here. The reason is that it
is not possible for $[\phi_{1},\dot{\phi}_{1}]$ to be a linear combination
of $\phi_{1}$ and $\dot{\phi}_{1}$, since the orders of spatial
derivatives of $\mathcal{F}_{1}$, $\partial_{i}\mathcal{F}^{i}$
and $\partial^{2}\mathcal{F}_{2}$ are different. This can be seen
more explicitly by evaluating the Poisson bracket $[\phi_{1}(\vec{x}),\dot{\phi}_{1}(\vec{y})]$
with arbitrary test functions $A(\vec{x})$ and $B(\vec{y})$, which
yields
\begin{align*}
0= & \int\mathrm{d}^{3}x\mathrm{d}^{3}yA(\vec{x})B(\vec{y})[\phi_{1}(\vec{x}),\dot{\phi}_{1}(\vec{y})]\\
\simeq & \int\mathrm{d}^{3}x\mathrm{d}^{3}y\left(A(\vec{x})B(\vec{y})\mathcal{F}_{1}(\vec{x})-A(\vec{x})\frac{\partial B(\vec{y})}{\partial y^{i}}\mathcal{F}^{i}(\vec{x})+A(\vec{x})\frac{\partial^{2}B(\vec{y})}{\partial y^{i}\partial y_{i}}\mathcal{F}_{2}(\vec{x})\right)\delta^{(3)}(y-\vec{x})\\
= & \int\mathrm{d}^{3}x\left(A(\vec{x})B(\vec{x})\mathcal{F}_{1}(\vec{x})-A(\vec{x})\frac{\partial B(\vec{x})}{\partial x^{i}}\mathcal{F}^{i}(\vec{x})+A(\vec{x})\frac{\partial^{2}B(\vec{x})}{\partial x^{i}\partial x_{i}}\mathcal{F}_{2}(\vec{x})\right).
\end{align*}
Due to the arbitrariness of the test functions, (\ref{firstdgc_xpl})
is satisfied if and only if all the three coefficients are vanishing
separately, i.e.,
\begin{equation}
\mathcal{F}_{1}=0,\quad\mathcal{F}^{i}=0,\quad\mathcal{F}_{2}=0.\label{dgc1de}
\end{equation}
Essentially, this is because they involve terms of differential orders
in spatial derivatives. 

(\ref{dgc1de}) are 3 differential equations for the coefficients,
which then constitute the explicit form of the first degeneracy condition.
After plugging (\ref{Pii}), i.e., $\varPi^{i}=2h_{4,1}\partial_{0}A^{i}+h_{4,3}\partial^{i}A_{0}$,
we can solve
\begin{equation}
f_{4,2}=C_{1}A_{0}+C_{2},\quad h_{4,1}=\frac{1}{C_{1}A_{0}+C_{2}},
\end{equation}
where $C_{1}$ and $C_{2}$ are constants. As we have mentioned before,
in this work we focus on the case in which the coefficients are polynomials
of $A_{0}$ and $\bar{X}$. This ansatz is adopted to make the solution
space of the (functional) degeneracy conditions explicitly classifiable
and to exhibit closed-form families of 2 DOF Lagrangians. The degeneracy
conditions themselves can be formulated for more general coefficient
functions (e.g. analytic functions admitting local series expansions
around backgrounds), and the Hamiltonian/constraint logic below does
not rely on polynomiality except when solving the conditions explicitly.
With this assumption, there is no other possibility than that $h_{4,1}$
is a constant. In other words, we must choose $C_{1}=0$ in the above
solutions. 

With this result, the Lagrangians satisfying the first degeneracy
condition can be specified explicitly. For $\mathcal{L}_{2}=f_{2}(A_{0},\bar{X})$,
we find it must be linear in $A_{0}$. That is, we may write
\begin{equation}
\mathcal{L}_{2}=f_{2}\equiv\bar{f}_{2}(\bar{X})+A_{0}\tilde{f}_{2}(\bar{X}),\label{Lag2_dg1}
\end{equation}
where $\bar{f}_{2}$ and $\tilde{f}_{2}$ are general polynomials
of $\bar{X}$. For $\mathcal{L}_{3}$, we find that $f_{3,1}$ cannot
depend on $\bar{X}$ but can be a general polynomial of $A_{0}$,
whereas $f_{3,2}$ must be linear in $A_{0}$. Thus we can write
\begin{equation}
\mathcal{L}_{3}=f_{3,1}(A_{0})\dot{A}_{0}+f_{3,2}(A_{0},\bar{X})\partial_{i}A^{i}\simeq\left(\bar{f}_{3,2}+A_{0}\tilde{f}_{3,2}\right)\partial_{i}A^{i},\label{Lag3_dg1}
\end{equation}
where $\bar{f}_{3,2}$ and $\tilde{f}_{3,2}$ are general polynomials
of $\bar{X}$. Note $\dot{A}_{0}$ term in $\mathcal{L}_{3}$ is removed
by integration by parts. For $\mathcal{L}_{4}$, we find that $h_{4,1}$,
$h_{4,2}$, $h_{4,3}$ and $f_{4,2}$ must be constants satisfying
\begin{equation}
f_{4,2}(f_{4,2}+h_{4,3})=4\left(h_{4,2}h_{4,1}-\frac{(h_{4,3})^{2}}{4}\right).\label{constrel}
\end{equation}
Thus we may denote
\begin{align}
h_{4,1} & =\alpha,\quad f_{4,2}=\beta,\quad h_{4,3}=\gamma,\\
h_{4,2} & =\frac{(\beta+\gamma)^{2}-\beta\gamma}{4\alpha}.
\end{align}
We also find that $f_{4,3}$, $f_{4,4}$ and $f_{4,5}$ are linear
in $A_{0}$, yielding
\begin{align}
\mathcal{L}_{4}\simeq & \left(\beta+\gamma\right)\dot{A}_{i}\partial^{i}A_{0}+\alpha\dot{A}_{i}\dot{A}^{i}+\frac{(\beta+\gamma)^{2}-\beta\gamma}{4\alpha}\partial_{i}A_{0}\partial^{i}A_{0}\nonumber \\
 & +\left(\bar{f}_{4,3}+A_{0}\tilde{f}_{4,3}\right)(\partial_{i}A^{i})^{2}+\left(\bar{f}_{4,4}+A_{0}\tilde{f}_{4,4}\right)\partial_{i}A_{j}\partial^{i}A^{j}+\left(\bar{f}_{4,5}+A_{0}\tilde{f}_{4,5}\right)\partial_{i}A_{j}\partial^{j}A^{i},\label{Lag4_dg1}
\end{align}
where $\bar{f}_{4,3}$, $\tilde{f}_{4,3}$, $\bar{f}_{4,4}$, $\tilde{f}_{4,4}$,
$\bar{f}_{4,5}$ and $\tilde{f}_{4,5}$ are general polynomials of
$\bar{X}$.

When the first degeneracy condition is satisfied, i.e., $[\phi_{1},\dot{\phi}_{1}]\approx0$,
we have
\begin{equation}
\mathcal{A}=\frac{1}{2c_{4,1}}(f_{4,2}-h_{4,3}),\quad\mathcal{B}=0,\quad\mathcal{C}_{i}^{(0,i)}=0,
\end{equation}
and
\begin{equation}
\mathcal{D}^{(0,i)}=-\frac{\partial}{\partial A_{0}}[f_{2}+f_{3,2}\partial_{i}A^{i}+f_{4,3}(\partial_{i}A^{i})^{2}+f_{4,4}\partial_{i}A_{j}\partial^{i}A^{j}+f_{4,5}\partial_{i}A_{j}\partial^{j}A^{i}],
\end{equation}
where the coefficients are given as in (\ref{Lag2_dg1}), (\ref{Lag3_dg1})
and (\ref{Lag4_dg1}). As a result, we can rewrite $\dot{\phi}$ as
$\dot{\phi}=\mathcal{A}\partial_{i}\varPi^{i}+\mathcal{D}^{(0,i)}$. 

In general, the next consistency condition $\ddot{\phi}_{1}\approx0$
gives rise to a tertiary constraint, since the Poisson bracket $[\dot{\phi}_{1}(\vec{x}),H_{\mathrm{C}}]$
does not vanish weakly. The Dirac matrix now takes the schematic form
\begin{equation}
B^{ab}(\vec{x},\vec{y})=\begin{array}{|c|ccc|}
\hline [\cdot,\cdot] & \phi_{1} & \dot{\phi}_{1} & \ddot{\phi}_{1}\\
\hline \phi_{1} & 0 & 0 & X\\
\dot{\phi}_{1} & 0 & X & X\\
\ddot{\phi}_{1} & X & X & X
\\\hline \end{array},\label{Bab}
\end{equation}
where again ``$X$'' denotes generally non-vanishing entries. If
no additional constraints appear, (\ref{Bab}) implies that all three
constraints $\{\phi_{1},\dot{\phi}_{1},\ddot{\phi}_{1}\}$ are second-class,
leading formally to 2.5 DOFs. This can be understood as follows. For
finite-dimensional systems, an odd-dimensional antisymmetric matrix
has vanishing determinant and hence cannot be invertible (nondegenerate).
However, this does not directly apply to $B_{ab}(\vec{x},\vec{y})$
in (\ref{Bab}). In field theory, the object $B_{ab}(\vec{x},\vec{y})$
in (\ref{Bab}) should be regarded as a differential operator acting
on test functions, rather than a finite-dimensional matrix. Indeed,
in field theories it is possible for an odd number of constraints
(per spatial point) to be genuinely second-class when the Poisson
brackets contain spatial derivatives of $\delta$-functions and the
corresponding operator is invertible after imposing appropriate boundary
conditions. A classical example is the Floreanini-Jackiw chiral boson,
which is commonly described as carrying one-half physical DOF \citep{Floreanini:1987as}
(see also \citep{Abreu:2004kn}). Similar phenomena have also been
discussed in Lorentz violating field theories, such as Hořava-Lifshitz
gravity \citep{Blas:2009yd} and spatially covariant gravity theories
\citep{Gao:2018znj,Gao:2019twq,Yao:2020tur,Iyonaga:2020bmm}, where
an additional condition is required in general to ensure that the
would-be half mode is absent and only the desired propagating modes
remain. With this understanding, (\ref{Bab}) indicates that if $B_{ab}(\vec{x},\vec{y})$
is invertible as an operator on the space of test functions, then
$\{\phi_{1},\dot{\phi}_{1},\ddot{\phi}_{1}\}$ are second-class and
the theory exhibits a residual half DOF, so that the total number
of DOFs is formally 2.5. Our goal is precisely to exclude this possibility
and ensure that only the two transverse polarizations propagate. 

To achieve this, the Dirac matrix $B^{ab}$ in (\ref{Bab}) must be
degenerate, i.e.,
\begin{equation}
\det B^{ab}(\vec{x},\vec{y})=0,\label{detBab_dg}
\end{equation}
which means there exists a nontrivial null vector $V_{b}(\vec{y})$
such that $\int\mathrm{d}^{3}y\,B^{ab}(\vec{x},\vec{y})V_{b}(\vec{y})=0$.
We therefore refer to (\ref{detBab_dg}) as the second degeneracy
condition. It can be shown that $\det B^{ab}$ is proportional to
the Poisson brackets $[\dot{\phi}_{1},\dot{\phi}_{1}]$ and $[\phi_{1},\ddot{\phi}_{1}]$,
while it has nothing to do with $[\dot{\phi}_{1},\ddot{\phi}_{1}]$
or $[\ddot{\phi}_{1},\ddot{\phi}_{1}]$. Hence the condition (\ref{detBab_dg})
can be realized with either of the following two choices:
\begin{equation}
[\dot{\phi}_{1}(\vec{x}),\dot{\phi}_{1}(\vec{y})]\approx0,\label{degcon2_1}
\end{equation}
or
\begin{equation}
[\phi_{1}(\vec{x}),\ddot{\phi}_{1}(\vec{y})]\approx0.\label{degcon2_2}
\end{equation}
If either (\ref{degcon2_1}) or (\ref{degcon2_2}) is satisfied, we
have $\det B^{ab}=0$. We refer to (\ref{degcon2_1}) and (\ref{degcon2_2})
as the two branches of the second degeneracy condition.

In the following section, we explore the conditions on the Lagrangians
under which either (\ref{degcon2_1}) or (\ref{degcon2_2}) is satisfied.
In certain cases, both of them can be satisfied simultaneously. As
we will see below, if one of the two branches of the second degeneracy
condition is satisfied, the theory propagates exactly two dynamical
DOFs.

\section{The second degeneracy condition \label{sec:dgncon2}}

In the previous section, we showed that to further eliminate the remaining
half DOF, the second degeneracy condition (\ref{detBab_dg}) must
be satisfied. As discussed, this requirement can be realized by either
(\ref{degcon2_1}) or (\ref{degcon2_2}), which we refer to as branch
1 and branch 2 of the second degeneracy condition. These two branches
correspond to distinct patterns of constraints and lead to different
classes of theories, which we analyze separately below.

\subsection{Branch 1 \label{subsec:branch1}}

We first consider branch 1, corresponding to (\ref{degcon2_1}). By
evaluating the Poisson bracket $[\dot{\phi}_{1},\dot{\phi}_{1}]$
explicitly, we obtain
\begin{align}
[\dot{\phi}_{1}(\vec{x}),\dot{\phi}_{1}(\vec{y})] & =\mathcal{F}^{*k}(\vec{x})\frac{\partial}{\partial y^{k}}\delta^{(3)}(\vec{y}-\vec{x})+\mathcal{F}^{*}(\vec{x})\frac{\partial^{2}}{\partial y^{k}\partial y_{k}}\delta^{(3)}(\vec{y}-\vec{x}),\label{PB_dfdf}
\end{align}
where
\begin{equation}
\mathcal{F}^{*k}=2\mathcal{A}\left[\frac{\partial\mathcal{D}^{(0,i)}}{\partial A_{k}}+\frac{\partial}{\partial A_{l}}\left(\frac{\partial f_{3,2}}{\partial A_{0}}\right)\partial^{k}A_{l}+2\frac{\partial}{\partial A_{l}}\left(\frac{\partial f_{4,4}}{\partial A_{0}}\right)\partial^{i}A^{k}\partial_{i}A_{l}\right],\label{calFsk}
\end{equation}
and
\begin{equation}
\mathcal{F}^{*}=2\mathcal{A}\left[\frac{\partial f_{3,2}}{\partial A_{0}}+2\frac{\partial}{\partial A_{0}}\left(f_{4,3}+f_{4,4}+f_{4,5}\right)\partial_{i}A^{i}\right].\label{calFs}
\end{equation}
Again, with the technique of test functions and the fact that $[\dot{\phi}_{1},\dot{\phi}_{1}]$
cannot be expressed as a linear combination of $\phi_{1}$ and $\dot{\phi}_{1}$,
the condition $[\dot{\phi}_{1}(\vec{x}),\dot{\phi}_{1}(\vec{y})]\approx0$
is equivalent to requiring
\begin{equation}
\mathcal{F}^{*k}=0,\quad\mathcal{F}^{*}=0.\label{degcon2_b1_sol}
\end{equation}
Depending on whether $\mathcal{A}=0$ or not, we have two cases of
solutions to (\ref{degcon2_b1_sol}), which we will discuss below.

\subsubsection{Case 1}

The simplest solution to (\ref{degcon2_b1_sol}) is 
\begin{equation}
\mathcal{A}=0,\label{degcon2_b1_s1}
\end{equation}
which implies
\begin{equation}
f_{4,2}=-h_{4,3}.\label{degcon2_b1_s1a}
\end{equation}

After some manipulations, we can determine the Lagrangians satisfying
(\ref{degcon2_b1_s1}) or equivalently (\ref{degcon2_b1_s1a}), together
with the first degeneracy condition. For $\mathcal{L}_{2}=f_{2}(A_{0},\bar{X})$,
we find it must be linear in $A_{0}$, i.e.,
\begin{equation}
\mathcal{L}_{2}=\bar{f}_{2}(\bar{X})+A_{0}\tilde{f}_{2}(\bar{X}),\label{Lag2_b1c1}
\end{equation}
where $\bar{f}_{2}$ and $\tilde{f}_{2}$ are general polynomials
of $\bar{X}$. For $\mathcal{L}_{3}$, we find that $f_{3,1}$ cannot
depend on $\bar{X}$ but can be a general polynomial of $A_{0}$,
and $f_{3,2}$ is linear in $A_{0}$. Thus we can write
\begin{equation}
\mathcal{L}_{3}=f_{3,1}\dot{A}_{0}+\left(\bar{f}_{3,2}+A_{0}\tilde{f}_{3.2}\right)\partial_{i}A^{i}\simeq\left(\bar{f}_{3,2}+A_{0}\tilde{f}_{3.2}\right)\partial_{i}A^{i},\label{Lag3_b1c1}
\end{equation}
where $\bar{f}_{3,2}$ and $\tilde{f}_{3,2}$ are general polynomials
of $\bar{X}$. In the above, we have used the fact that $f_{3,1}(A_{0})\dot{A}_{0}$
can be reduced by integration by parts. For $\mathcal{L}_{4}$, we
find that $h_{4,1}$, $h_{4,2}$, $h_{4,3}$ and $f_{4,2}$ must be
constants satisfying 
\begin{equation}
-h_{4,3}=f_{4,2}=2\sqrt{h_{4,2}h_{4,1}}.
\end{equation}
Defining
\begin{equation}
h_{4,1}=\alpha_{1},\quad f_{4,2}=-h_{4,3}=\beta_{1},\quad h_{4,2}=\frac{\beta_{1}^{2}}{4\alpha_{1}},
\end{equation}
and noting that $f_{4,3}$, $f_{4,4}$, and $f_{4,5}$ are linear
in $A_{0}$, we finally obtain 
\begin{align}
\mathcal{L}_{4}= & \alpha_{1}\dot{A}_{i}\dot{A}^{i}+\frac{\beta_{1}^{2}}{4\alpha_{1}}\partial_{i}A_{0}\partial^{i}A_{0}+\left(\bar{f}_{4,3}+A_{0}\tilde{f}_{4,3}\right)(\partial_{i}A^{i})^{2}\nonumber \\
 & +\left(\bar{f}_{4,4}+A_{0}\tilde{f}_{4,4}\right)\partial_{i}A_{j}\partial^{i}A^{j}+\left(\bar{f}_{4,5}+A_{0}\tilde{f}_{4,5}\right)\partial_{i}A_{j}\partial^{j}A^{i},\label{Lag4_b1c1}
\end{align}
where $\bar{f}_{4,3}$, $\tilde{f}_{4,3}$, $\bar{f}_{4,4}$, $\tilde{f}_{4,4}$,
$\bar{f}_{4,5}$ and $\tilde{f}_{4,5}$ are general polynomials of
$\bar{X}$. 

With the above solutions for the Lagrangians, the Dirac matrix takes
the form
\begin{equation}
\bar{B}^{ab}(\vec{x},\vec{y})=\begin{array}{|c|ccc|}
\hline [\cdot,\cdot] & \phi_{1} & \dot{\phi}_{1} & \ddot{\phi}_{1}\\
\hline \phi_{1} & 0 & 0 & X\\
\dot{\phi}_{1} & 0 & 0 & X\\
\ddot{\phi}_{1} & X & X & X
\\\hline \end{array},
\end{equation}
which is degenerate as $\det\bar{B}^{ab}(\vec{x},\vec{y})=0$. This
implies the existence of one first-class constraint that can be expressed
as a linear combination of the three constraints $\phi^{a}\equiv\{\phi_{1},\dot{\phi}_{1},\ddot{\phi}_{1}\}$.
This can be seen from the fact that the degeneracy of $B^{ab}$ implies
the existence of a null eigenvector
\begin{equation}
\int\mathrm{d}^{3}y\,\bar{B}^{ab}(\vec{x},\vec{y})\mathcal{V}_{b}(\vec{y})=0.
\end{equation}
Using this null eigenvector, one can build a new constraint by combining
the constraints
\begin{equation}
\varPhi=\int\mathrm{d}^{3}x\,\phi^{a}(\vec{x})\mathcal{V}_{a}(\vec{x}),
\end{equation}
which is clearly first-class as $\left[\varPhi,\phi^{a}(\vec{x})\right]=0$.

For our purposes, we do not need the explicit expression for the new
first-class constraint. Instead, we formally denote it as $\tilde{\phi}_{1}$
and the rest two newly combined constraints as $\tilde{\phi}_{2}$
and $\tilde{\phi}_{3}$. Hence the new Dirac matrix becomes
\[
\tilde{B}^{ab}(\vec{x},\vec{y})=\begin{array}{|c|ccc|}
\hline [\cdot,\cdot] & \tilde{\phi}_{1} & \tilde{\phi}_{2} & \tilde{\phi}_{3}\\
\hline \tilde{\phi}_{1} & 0 & 0 & 0\\
\tilde{\phi}_{2} & 0 & X & X\\
\tilde{\phi}_{3} & 0 & X & X
\\\hline \end{array},
\]
where $\tilde{\phi}_{a}$ are linear combinations of $\phi_{1},\dot{\phi}_{1},\ddot{\phi}_{1}$.
Generally, the Poisson bracket $[\phi_{1}(\vec{x}),\ddot{\phi}_{1}(\vec{y})]$
does not vanish. Therefore the consistency condition for $\ddot{\phi}_{1}$
does not yield any further constraint. As a result, the theory possesses
three constraints, of which one is first-class, two are second-class.
For later convenience, we refer to theories in this case as ``type-I''
theories. The number of DOFs is counted as
\begin{equation}
\#_{\mathrm{dof}}=\frac{2\times4_{\mathrm{var}}-2\times1_{1\mathrm{st}}-1\times2_{2\mathrm{st}}}{2}=2.
\end{equation}
As we can see, the number of DOFs of the theory does reduce to two
in this case. 

\subsubsection{Case 2 \label{subsec:b1c2}}

Let us now focus on the other case, i.e., when $\mathcal{A}\neq0$.
In this case, the second degeneracy condition (\ref{degcon2_b1_sol})
implies that quantities in the square brackets in (\ref{calFsk})
and (\ref{calFs}) must be vanishing. This case seems more complicated,
especially when explicitly substituting conjugate momenta results
in higher order differential equations. 

Nevertheless, after some manipulations, we can determine the Lagrangians
satisfying both the first and the second degeneracy conditions in
this case. For $\mathcal{L}_{2}=f_{2}(\bar{X})$, it must be linear
in $\bar{X}$ and independent of $A_{0}$. That is, we may write
\begin{equation}
\mathcal{L}_{2}=\bar{f}_{2}\bar{X}+\tilde{f}_{2},\label{Lag2_b1c2}
\end{equation}
where $\bar{f}_{2}$ and $\tilde{f}_{2}$ are constants. For $\mathcal{L}_{3}$,
$f_{3,1}$ is again independent of $\bar{X}$ but can generally depend
on $A_{0}$, while $f_{3,2}$ is linear in $\bar{X}$. Thus we can
write
\begin{equation}
\mathcal{L}_{3}=f_{3,1}\dot{A}_{0}+\left(\bar{f}_{3,2}\bar{X}+\tilde{f}_{3.2}\right)\partial_{i}A^{i}\simeq\left(\bar{f}_{3,2}\bar{X}+\tilde{f}_{3,2}\right)\partial_{i}A^{i},\label{Lag3_b1c2}
\end{equation}
where $\bar{f}_{3,2}$ and $\tilde{f}_{3,2}$ are constants and again
we used integration by parts to suppress the term involving $\dot{A}_{0}$.
For $\mathcal{L}_{4}$, the coefficients $h_{4,1}$, $h_{4,2}$, $h_{4,3}$
and $f_{4,2}$ must be constants satisfying 
\begin{equation}
f_{4,2}(f_{4,2}+h_{4,3})=4\left(h_{4,2}h_{4,1}-\frac{(h_{4,3})^{2}}{4}\right),
\end{equation}
which can be parametrized as 
\begin{align}
h_{4,1} & =\alpha_{2},\quad f_{4,2}=\beta_{2},\quad h_{4,3}=\gamma_{2},\\
h_{4,2} & =\frac{(\beta_{2}+\gamma_{2})^{2}-\beta_{2}\gamma_{2}}{4\alpha_{2}}.
\end{align}
With $f_{4,3}$, $f_{4,4}$, and $f_{4,5}$ linear in $\bar{X}$,
the final form of $\mathcal{L}_{4}$ is
\begin{align}
\mathcal{L}_{4}\simeq & \left(\beta_{2}+\gamma_{2}\right)\dot{A}_{i}\partial^{i}A_{0}+\alpha_{2}\dot{A}_{i}\dot{A}^{i}+\frac{(\beta_{2}+\gamma_{2})^{2}-\beta_{2}\gamma_{2}}{4\alpha_{2}}\partial_{i}A_{0}\partial^{i}A_{0}\nonumber \\
 & +\left(\bar{f}_{4,3}\bar{X}+\tilde{f}_{4,3}\right)(\partial_{i}A^{i})^{2}+\left(\bar{f}_{4,4}\bar{X}+\tilde{f}_{4,4}\right)\partial_{i}A_{j}\partial^{i}A^{j}+\left(\bar{f}_{4,5}\bar{X}+\tilde{f}_{4,5}\right)\partial_{i}A_{j}\partial^{j}A^{i},\label{Lag4_b1c2}
\end{align}
where $\bar{f}_{4,3}$, $\tilde{f}_{4,3}$, $\bar{f}_{4,4}$, $\tilde{f}_{4,4}$,
$\bar{f}_{4,5}$ and $\tilde{f}_{4,5}$ are constants. 

The Poisson bracket $[\phi_{1}(\vec{x}),\ddot{\phi}_{1}(\vec{y})]$
vanishes automatically, yielding the Dirac matrix 
\begin{equation}
B^{ab}(\vec{x},\vec{y})=\begin{array}{|c|ccc|}
\hline [\cdot,\cdot] & \phi_{1} & \dot{\phi}_{1} & \ddot{\phi}_{1}\\
\hline \phi_{1} & 0 & 0 & 0\\
\dot{\phi}_{1} & 0 & 0 & X\\
\ddot{\phi}_{1} & 0 & X & X
\\\hline \end{array}.
\end{equation}
We find that the Poisson bracket $[\ddot{\phi}_{1}(\vec{x}),H_{\mathrm{C}}]$
is vanishing on the constraint surface, and
\[
\dddot{\phi}_{1}\sim-\frac{\beta_{2}+\gamma_{2}}{\alpha_{2}}\partial^{3}(f_{4,3}+f_{4,4}+f_{4,5})=0,
\]
which implies that there is no quaternary constraint. Then the number
of DOF is counted as
\begin{equation}
\#_{\mathrm{DOF}}=\frac{2\times4_{\mathrm{var}}-2\times1_{1\mathrm{st}}-1\times2_{2\mathrm{nd}}}{2}=2.
\end{equation}
Therefore, theories in this case is also of ``type-I''. 

Interestingly, as we will see in Sec. \ref{subsec:b2c1}, the case
discussed here can be also viewed as a special solution in branch
2.

\subsection{Branch 2 \label{subsec:branch2}}

We now consider branch 2, corresponding to (\ref{degcon2_2}).

To proceed, we require the explicit expression for the tertiary constraint
$\ddot{\phi}_{1}$, obtained from the consistency condition of $\dot{\phi}_{1}\approx0$.
Assuming that the Poisson bracket $[\dot{\phi}_{1}(\vec{x}),H_{\mathrm{C}}]$
does not vanish weakly, we find 
\begin{align}
\ddot{\phi}_{1} & =\mathcal{C}_{k}^{*(0,i)}\varPi^{k}+\mathcal{D}^{*(0,i)},
\end{align}
where the coefficients $\mathcal{C}^{*k(0,i)}$ and $\mathcal{D}^{*(0,i)}$
contain no momenta and are given by
\begin{align}
\mathcal{C}^{*k(0,i)}= & \bigg[\frac{\partial\mathcal{D}^{(0,i)}}{\partial A_{k}}+\frac{\partial}{\partial A_{l}}\left(\frac{\partial f_{3,2}}{\partial A_{0}}+2\frac{\partial f_{4,3}}{\partial A_{0}}\partial_{i}A^{i}\right)\partial^{k}A_{l}\nonumber \\
 & +2\frac{\partial}{\partial A_{l}}\left(\frac{\partial f_{4,4}}{\partial A_{0}}\partial^{i}A^{k}+\frac{\partial f_{4,5}}{\partial A_{0}}\partial^{k}A^{i}\right)\partial_{i}A_{l}\bigg]\mathcal{K},\label{calC0i_ast}
\end{align}
with $\mathcal{K}$ given in (\ref{calK}), $\mathcal{D}^{(0,i)}$
given in (\ref{calD0i}), and
\begin{align}
\mathcal{D}^{*(0,i)}= & \left[\frac{\partial\mathcal{D}^{(0,i)}}{\partial A_{k}}+\frac{\partial}{\partial A_{l}}\left(\frac{\partial f_{3,2}}{\partial A_{0}}+2\frac{\partial f_{4,3}}{\partial A_{0}}\partial_{i}A^{i}\right)\partial^{k}A_{l}+2\frac{\partial}{\partial A_{l}}\left(\frac{\partial f_{4,4}}{\partial A_{0}}\partial^{i}A^{k}+\frac{\partial f_{4,5}}{\partial A_{0}}\partial^{k}A^{i}\right)\partial_{i}A_{l}\right]\mathcal{W}_{k}^{(0)}\nonumber \\
 & +\mathcal{A}\bigg[\frac{\partial^{2}\mathcal{V}^{(0,i)}}{\partial A_{0}\partial A_{k}}\partial_{k}A_{0}+\mathcal{A}\frac{\partial^{2}\mathcal{V}^{(0,i)}}{\partial A_{l}\partial A_{k}}\partial_{k}A_{l}\nonumber \\
 & \quad+\frac{\partial}{\partial A_{0}}\left(\frac{\partial f_{3,2}}{\partial A_{l}}\right)\partial^{k}A_{l}\partial_{k}A_{0}+\frac{\partial}{\partial A_{m}}\left(\frac{\partial f_{3,2}}{\partial A_{0}}\partial^{k}A_{0}+\frac{\partial f_{3,2}}{\partial A_{l}}\partial^{k}A_{l}\right)\partial_{k}A_{m}\nonumber \\
 & \quad+\frac{\partial}{\partial A_{0}}\left(\frac{\partial f_{4,3}}{\partial A_{l}}\right)\partial_{i}A^{i}\partial^{k}A_{l}\partial_{k}A_{0}+\frac{\partial}{\partial A_{m}}\left(\frac{\partial f_{4,3}}{\partial A_{0}}\partial_{i}A^{i}\partial^{k}A_{0}+\frac{\partial f_{4,3}}{\partial A_{l}}\partial_{i}A^{i}\partial^{k}A_{l}\right)\partial_{k}A_{m}\nonumber \\
 & \quad+\frac{\partial}{\partial A_{0}}\left(\frac{\partial f_{4,4}}{\partial A_{l}}\right)\partial^{i}A^{k}\partial_{i}A_{l}\partial_{k}A_{0}+\frac{\partial}{\partial A_{m}}\left(\frac{\partial f_{4,4}}{\partial A_{0}}\partial^{i}A^{k}\partial_{i}A_{0}+\frac{\partial f_{4,4}}{\partial A_{l}}\partial^{i}A^{k}\partial_{i}A_{l}\right)\partial_{k}A_{m}\nonumber \\
 & \quad+\frac{\partial}{\partial A_{0}}\left(\frac{\partial f_{4,5}}{\partial A_{l}}\right)\partial^{k}A^{i}\partial_{i}A_{l}\partial_{k}A_{0}+\frac{\partial}{\partial A_{m}}\left(\frac{\partial f_{4,5}}{\partial A_{0}}\partial^{k}A^{i}\partial_{i}A_{0}+\frac{\partial f_{4,5}}{\partial A_{l}}\partial^{k}A^{i}\partial_{i}A_{l}\right)\partial_{k}A_{m}\bigg],\label{calD0i_ast}
\end{align}
where $\mathcal{A}$ is given in (\ref{calA}), $\mathcal{W}_{k}^{(0)}$
in (\ref{calW0}), and $\mathcal{V}^{(0,i)}$ in (\ref{calV0i}).

After a tedious but straightforward calculation, we obtain the Poisson
bracket 
\begin{align}
[\phi_{1}(\vec{x}),\ddot{\phi}_{1}(\vec{y})] & =\tilde{\mathcal{F}}^{k}\frac{\partial}{\partial y^{k}}\delta^{(3)}(\vec{y}-\vec{x})+\tilde{\mathcal{F}}_{1}\delta^{(3)}(\vec{y}-\vec{x}),
\end{align}
where
\begin{equation}
\tilde{\mathcal{F}}_{1}=-\frac{\partial\mathcal{C}^{*k(0,i)}}{\partial A_{0}}\varPi_{k}-\frac{\partial\mathcal{D}^{*(0,i)}}{\partial A_{0}}-\frac{\partial f_{4,2}}{\partial A_{k}}\partial_{i}A^{i}\mathcal{C}_{k}^{*(0,i)}-\frac{\partial f_{3,1}}{\partial A_{k}}\mathcal{C}_{k}^{*(0,i)},
\end{equation}
and
\begin{align}
\tilde{\mathcal{F}}^{k}= & \left[\frac{\partial\mathcal{D}^{(0,i)}}{\partial A_{k}}+\frac{\partial}{\partial A_{l}}\left(\frac{\partial f_{3,2}}{\partial A_{0}}+2\frac{\partial f_{4,3}}{\partial A_{0}}\partial_{i}A^{i}\right)\partial^{k}A_{l}+2\frac{\partial}{\partial A_{l}}\left(\frac{\partial f_{4,4}}{\partial A_{0}}\partial^{i}A^{k}+\frac{\partial f_{4,5}}{\partial A_{0}}\partial^{k}A^{i}\right)\partial_{i}A_{l}\right]\frac{h_{4,3}}{2h_{4,1}}-f_{4,2}\mathcal{C}^{*k(0,i)}\nonumber \\
 & +\mathcal{A}\bigg[\frac{\partial^{2}\mathcal{V}^{(0,i)}}{\partial A_{0}\partial A_{k}}+\frac{\partial}{\partial A_{0}}\left(\frac{\partial f_{3,2}}{\partial A_{l}}\right)\partial^{k}A_{l}+\frac{\partial}{\partial A_{m}}\left(\frac{\partial f_{3,2}}{\partial A_{0}}\right)\partial^{k}A_{m}+\frac{\partial}{\partial A_{0}}\left(\frac{\partial f_{4,3}}{\partial A_{l}}\right)\partial_{i}A^{i}\partial^{k}A_{l}\nonumber \\
 & \quad+\frac{\partial}{\partial A_{m}}\left(\frac{\partial f_{4,3}}{\partial A_{0}}\right)\partial_{i}A^{i}\partial^{k}A_{m}+\frac{\partial}{\partial A_{0}}\left(\frac{\partial f_{4,4}}{\partial A_{l}}\right)\partial^{i}A^{k}\partial_{i}A_{l}+\frac{\partial}{\partial A_{m}}\left(\frac{\partial f_{4,4}}{\partial A_{0}}\right)\partial^{k}A^{i}\partial_{i}A_{m}\nonumber \\
 & \quad+\frac{\partial}{\partial A_{0}}\left(\frac{\partial f_{4,5}}{\partial A_{l}}\right)\partial^{k}A^{i}\partial_{i}A_{l}+\frac{\partial}{\partial A_{m}}\left(\frac{\partial f_{4,5}}{\partial A_{0}}\right)\partial^{i}A^{k}\partial_{i}A_{m}\bigg].
\end{align}
Again, using the technique of test functions and noting that $[\phi_{1}(\vec{x}),\ddot{\phi}_{1}(\vec{y})]$
cannot be expressed as a linear combination of $\phi_{1}$, $\dot{\phi}_{1}$
and $\ddot{\phi}_{1}$, the condition $[\phi_{1}(\vec{x}),\ddot{\phi}_{1}(\vec{y})]\approx0$
implies
\begin{equation}
\tilde{\mathcal{F}}_{1}=0,\quad\tilde{\mathcal{F}}^{k}=0.\label{degcon2_b2_sol}
\end{equation}
By solving (\ref{degcon2_b2_sol}) with the substitution of $\varPi^{i}=2h_{4,1}\partial_{0}A^{i}+h_{4,3}\partial^{i}A_{0}$,
we can get the conditions on the coefficients. 

Solving (\ref{degcon2_b2_sol}), which consists of two differential
equations for the coefficients, is generally difficult. Nevertheless,
we can discuss two special cases by assuming that the coefficients
depend either on $A_{0}$ or $\bar{X}$ only, which simplifies the
analysis. As we will see below, one case yields four second-class
constraints, while the other case yields two first-class constraints. 

\subsubsection{Case 1 \label{subsec:b2c1}}

Let us focus on the first case, i.e., 
\begin{equation}
\frac{\partial f_{2}}{\partial A_{0}}=0,\quad\frac{\partial f_{m,n}}{\partial A_{0}}=0,
\end{equation}
which implies that the coefficients depend on $\bar{X}$ only. Under
this assumption, $\mathcal{C}^{*k(0,i)}$ and $\mathcal{D}^{*(0,i)}$
defined in (\ref{calC0i_ast}) and (\ref{calD0i_ast}) can be simplified
to be
\begin{equation}
\mathcal{C}^{*k(0,i)}=0,
\end{equation}
and
\begin{align}
\mathcal{D}^{*(0,i)}= & \mathcal{A}\bigg(\frac{\partial^{2}\mathcal{V}^{(0,i)}}{\partial A_{l}\partial A_{k}}\partial_{k}A_{l}+\frac{\partial^{2}f_{3,2}}{\partial A_{m}\partial A_{l}}\partial^{k}A_{l}\partial_{k}A_{m}+\frac{\partial^{2}f_{4,3}}{\partial A_{m}\partial A_{l}}\partial_{i}A^{i}\partial^{k}A_{l}\partial_{k}A_{m}\nonumber \\
 & +\frac{\partial^{2}f_{4,4}}{\partial A_{m}\partial A_{l}}\partial^{i}A^{k}\partial_{i}A_{l}\partial_{k}A_{m}+\frac{\partial^{2}f_{4,5}}{\partial A_{m}\partial A_{l}}\partial^{k}A^{i}\partial_{i}A_{l}\partial_{k}A_{m}\bigg).
\end{align}

In this case, the Dirac matrix takes the form
\begin{equation}
C^{ab}(\vec{x},\vec{y})=\begin{array}{|c|ccc|}
\hline [\cdot,\cdot] & \phi_{1} & \dot{\phi}_{1} & \ddot{\phi}_{1}\\
\hline \phi_{1} & 0 & 0 & 0\\
\dot{\phi}_{1} & 0 & X & X\\
\ddot{\phi}_{1} & 0 & X & X
\\\hline \end{array}.
\end{equation}
The consistency condition of $\ddot{\phi}_{1}$ yields
\begin{equation}
\int\mathrm{d}^{3}y\,[\phi_{1}(\vec{x}),\ddot{\phi}_{1}(\vec{y})]\nu(\vec{y})\approx-[\ddot{\phi}_{1}(\vec{x}),H_{\mathrm{C}}].\label{conscon_ddf}
\end{equation}
Here $\nu(\vec{y})$ is the Lagrange multiplier associated with the
primary constraint $\phi_{1}$ in the total Hamiltonian, which reads
$H_{\mathrm{T}}=H_{\mathrm{C}}+\int\mathrm{d}^{3}y\,\nu(\vec{y})\phi_{1}(\vec{y})$.
We need to calculate the Poisson bracket $[\ddot{\phi}_{1}(\vec{x}),H_{\mathrm{C}}]$
with $\ddot{\phi}_{1}=\mathcal{C}_{k}^{*(0,i)}\varPi^{k}+\mathcal{D}^{*(0,i)}$.
We find
\begin{align}
[\ddot{\phi}_{1}(\vec{x}),H_{\mathrm{C}}] & =\varTheta_{m}^{(0,i)}\varPi^{m}+\varDelta^{(0,i)},
\end{align}
where
\begin{align}
\varTheta^{m(0,i)}= & \mathcal{A}\mathcal{K}\Bigg[\frac{1}{\mathcal{A}}\frac{\partial\mathcal{D}^{*(0,i)}}{\partial A_{m}}-\partial_{k}\left(\frac{\partial^{2}\mathcal{V}^{(0,i)}}{\partial A_{m}\partial A_{k}}\right)+\partial^{m}\left(\frac{\partial^{2}f_{3,2}}{\partial A_{l}\partial A_{k}}\right)\partial_{k}A_{l}\nonumber \\
 & +2\partial^{m}\left(\frac{\partial^{2}f_{4,3}}{\partial A_{l}\partial A_{k}}\right)\partial_{i}A^{i}\partial_{k}A_{l}+2\partial_{i}\left(\frac{\partial^{2}f_{4,4}}{\partial A_{l}\partial A_{k}}\right)\partial^{i}A^{m}\partial_{k}A_{l}+2\partial^{m}\left(\frac{\partial^{2}f_{4,5}}{\partial A_{l}\partial A_{k}}\right)\partial^{m}A^{i}\partial_{k}A_{l}\nonumber \\
 & +\partial^{m}\left(\frac{\partial^{2}f_{4,3}}{\partial A_{i}\partial A_{l}}\right)\partial^{k}A_{l}\partial_{k}A_{i}+2\partial^{k}\left(\frac{\partial^{2}f_{4,3}}{\partial A_{l}\partial A_{m}}\right)\partial_{i}A^{i}\partial_{k}A_{l}\nonumber \\
 & +\partial^{i}\left(\frac{\partial^{2}f_{4,4}}{\partial A_{k}\partial A_{l}}\right)\partial_{i}A_{l}\partial^{m}A_{k}+\partial_{i}\left(\frac{\partial^{2}f_{4,4}}{\partial A_{l}\partial A_{m}}\right)\partial^{i}A^{k}\partial_{k}A_{l}+\partial_{k}\left(\frac{\partial^{2}f_{4,4}}{\partial A_{m}\partial A_{l}}\right)\partial^{i}A^{k}\partial_{i}A_{l}\nonumber \\
 & +\partial^{k}\left(\frac{\partial^{2}f_{4,5}}{\partial A_{i}\partial A_{l}}\right)\partial^{m}A_{l}\partial_{k}A_{i}+\partial_{i}\left(\frac{\partial^{2}f_{4,5}}{\partial A_{l}\partial A_{m}}\right)\partial^{k}A^{i}\partial_{k}A_{l}+\partial_{k}\left(\frac{\partial^{2}f_{4,5}}{\partial A_{m}\partial A_{l}}\right)\partial^{k}A^{i}\partial_{i}A_{l}\Bigg],
\end{align}
and
\begin{align}
\Delta^{(0,i)}= & \mathcal{A}\mathcal{W}_{m}^{(0)}\Bigg[\frac{1}{\mathcal{A}}\frac{\partial\mathcal{D}^{*(0,i)}}{\partial A_{m}}-\partial_{k}\left(\frac{\partial^{2}\mathcal{V}^{(0,i)}}{\partial A_{m}\partial A_{k}}\right)+\partial^{m}\left(\frac{\partial^{2}f_{3,2}}{\partial A_{l}\partial A_{k}}\right)\partial_{k}A_{l}\nonumber \\
 & +2\partial^{m}\left(\frac{\partial^{2}f_{4,3}}{\partial A_{l}\partial A_{k}}\right)\partial_{i}A^{i}\partial_{k}A_{l}+2\partial_{i}\left(\frac{\partial^{2}f_{4,4}}{\partial A_{l}\partial A_{k}}\right)\partial^{i}A^{m}\partial_{k}A_{l}+2\partial^{m}\left(\frac{\partial^{2}f_{4,5}}{\partial A_{l}\partial A_{k}}\right)\partial^{m}A^{i}\partial_{k}A_{l}\nonumber \\
 & +\partial^{m}\left(\frac{\partial^{2}f_{4,3}}{\partial A_{i}\partial A_{l}}\right)\partial^{k}A_{l}\partial_{k}A_{i}+2\partial^{k}\left(\frac{\partial^{2}f_{4,3}}{\partial A_{l}\partial A_{m}}\right)\partial_{i}A^{i}\partial_{k}A_{l}\nonumber \\
 & +\partial^{i}\left(\frac{\partial^{2}f_{4,4}}{\partial A_{k}\partial A_{l}}\right)\partial_{i}A_{l}\partial^{m}A_{k}+\partial_{i}\left(\frac{\partial^{2}f_{4,4}}{\partial A_{l}\partial A_{m}}\right)\partial^{i}A^{k}\partial_{k}A_{l}+\partial_{k}\left(\frac{\partial^{2}f_{4,4}}{\partial A_{m}\partial A_{l}}\right)\partial^{i}A^{k}\partial_{i}A_{l}\nonumber \\
 & +\partial^{k}\left(\frac{\partial^{2}f_{4,5}}{\partial A_{i}\partial A_{l}}\right)\partial^{m}A_{l}\partial_{k}A_{i}+\partial_{i}\left(\frac{\partial^{2}f_{4,5}}{\partial A_{l}\partial A_{m}}\right)\partial^{k}A^{i}\partial_{k}A_{l}+\partial_{k}\left(\frac{\partial^{2}f_{4,5}}{\partial A_{m}\partial A_{l}}\right)\partial^{k}A^{i}\partial_{i}A_{l}\Bigg].
\end{align}
Note that the coefficients have been restricted by the condition $[\phi_{1},\ddot{\phi}_{1}]\approx0$.
In evaluating the spatial derivatives above, only spatial derivatives
of $A_{i}$ contribute. For example,
\[
\partial^{m}\left(\frac{\partial^{2}f_{3,2}}{\partial A_{l}\partial A_{k}}\right)=\frac{\partial^{3}f_{3,2}}{\partial A_{n}\partial A_{l}\partial A_{k}}\partial^{m}A_{n}.
\]

Depending on whether $[\ddot{\phi}_{1}(\vec{x}),H_{\mathrm{C}}]$
is vanishing weakly or not, we can further consider two sub-cases.
In the special case where $[\ddot{\phi}_{1}(\vec{x}),H_{\mathrm{C}}]\approx0$
on the constraint surface, no quaternary constraint arises. In this
case, the Dirac matrix becomes
\begin{equation}
\bar{C}^{ab}(\vec{x},\vec{y})=\begin{array}{|c|ccc|}
\hline [\cdot,\cdot] & \phi_{1} & \dot{\phi}_{1} & \ddot{\phi}_{1}\\
\hline \phi_{1} & 0 & 0 & 0\\
\dot{\phi}_{1} & 0 & 0 & X\\
\ddot{\phi}_{1} & 0 & X & X
\\\hline \end{array},
\end{equation}
which indicates the emergence of a first-class constraint and thus
the enhancement of the gauge symmetry. The resulting theory possesses
one first-class constraint and two second-class constraints, and thus
belongs to ``type-I'' theories. Comparing with the case in Sec.
\ref{subsec:b1c2}, if the conditions discussed in \ref{subsec:b1c2}
are satisfied, the Poisson brackets $[\phi_{1},\ddot{\phi}_{1}]\approx0$
and $[\ddot{\phi}_{1},H_{\mathrm{C}}]\approx0$ will automatically
be satisfied, but not vice versa. In other words, the case 2 of branch
1 discussed in \ref{subsec:b1c2} is not independent in the sense
that it can be viewed as the sub-case discussed here, i.e., as the
special solution in case 1 of branch 2. 

We then turn to the general case, where $[\ddot{\phi}_{1}(\vec{x}),H_{\mathrm{C}}]$
does not vanish weakly. This leads to a new quaternary constraint,
yielding the Dirac matrix
\begin{equation}
D^{ab}(\vec{x},\vec{y})=\begin{array}{|c|cccc|}
\hline [\cdot,\cdot] & \phi_{1} & \dot{\phi}_{1} & \ddot{\phi}_{1} & \dddot{\phi}_{1}\\
\hline \phi_{1} & 0 & 0 & 0 & X\\
\dot{\phi}_{1} & 0 & X & X & X\\
\ddot{\phi}_{1} & 0 & X & X & X\\
\dddot{\phi}_{1} & X & X & X & X
\\\hline \end{array}.
\end{equation}
Generally, the Poisson bracket
\begin{align}
[\phi_{1}(\vec{x}),\dddot{\phi}_{1}(\vec{y})] & =\tilde{\mathcal{F}}_{k}^{*}\frac{\partial}{\partial y_{k}}\delta^{(3)}(\vec{y}-\vec{x})\label{PBfd3f}
\end{align}
with
\begin{equation}
\tilde{\mathcal{F}}_{k}^{*}=-f_{4,2}\varTheta_{k}^{(0,i)}\label{calFtldast}
\end{equation}
does not vanish weakly. Thus, in this case, the theory contains four
second-class constraints, leading to the number of DOFs
\begin{equation}
\#_{\mathrm{DOF}}=\frac{2\times4_{\mathrm{var}}-1\times4_{2\mathrm{nd}}}{2}=2.
\end{equation}
We refer to theories in this case as ``type-II'' theories. It is
clear that this is a new situation different from the previous ones. 

In this case, we can determine the Lagrangians satisfying the second
degeneracy condition. For $\mathcal{L}_{2}=f_{2}(\bar{X})$, we find
that it must be nonlinear in $\bar{X}$. Hence, we may write
\begin{equation}
\mathcal{L}_{2}=f_{2},\label{Lag2_b2c1}
\end{equation}
where $f_{2}$ is a nonlinear polynomial of $\bar{X}$. For $\mathcal{L}_{3}$,
we find that $f_{3,1}$ cannot depend on $\bar{X}$ but can be a general
polynomial of $A_{0}$, whereas $f_{3,2}$ must be nonlinear in $\bar{X}$.
Thus we can write 
\begin{equation}
\mathcal{L}_{3}=f_{3,1}(A_{0})\dot{A}_{0}+\bar{f}_{3,2}(\bar{X})\partial_{i}A^{i}\simeq\bar{f}_{3,2}(\bar{X})\partial_{i}A^{i},\label{Lag3_b2c1}
\end{equation}
where $\bar{f}_{3,2}$ is a nonlinear polynomial of $\bar{X}$. For
$\mathcal{L}_{4}$, we find that $h_{4,1}$, $h_{4,2}$, $h_{4,3}$
and $f_{4,2}$ must be constants satisfying 
\begin{equation}
f_{4,2}(f_{4,2}+h_{4,3})=4(h_{4,2}h_{4,1}-\frac{(h_{4,3})^{2}}{4}).
\end{equation}
We therefore denote
\begin{align}
h_{4,1} & =\alpha_{2},\quad f_{4,2}=\beta_{2},\quad h_{4,3}=\gamma_{2},\\
h_{4,2} & =\frac{(\beta_{2}+\gamma_{2})^{2}-\beta_{2}\gamma_{2}}{4\alpha_{2}}.
\end{align}
We also find that $f_{4,3}$, $f_{4,4}$, and $f_{4,5}$ are nonlinear
in $\bar{X}$. Thus finally we have
\begin{align}
\mathcal{L}_{4}\simeq & \left(\beta_{2}+\gamma_{2}\right)\dot{A}_{i}\partial^{i}A_{0}+\alpha_{2}\dot{A}_{i}\dot{A}^{i}+\frac{(\beta_{2}+\gamma_{2})^{2}-\beta_{2}\gamma_{2}}{4\alpha_{2}}\partial_{i}A_{0}\partial^{i}A_{0}\nonumber \\
 & +f_{4,3}\left(\partial_{i}A^{i}\right)^{2}+f_{4,4}\partial_{i}A_{j}\partial^{i}A^{j}+f_{4,5}\partial_{i}A_{j}\partial^{j}A^{i},\label{Lag4_b2c1}
\end{align}
where $f_{4,3}$, $f_{4,4}$, and $f_{4,5}$ are nonlinear polynomials
of $\bar{X}$. 

\subsubsection{Case 2}

Now let us turn to the other choice, given by
\begin{equation}
\frac{\partial f_{2}}{\partial A_{k}}=0,\;\frac{\partial f_{m,n}}{\partial A_{k}}=0,\label{b2c2_asp}
\end{equation}
which indicates that all the coefficients depend only on $A_{0}$.
In this case, $\mathcal{C}^{*k(0,i)}$ and $\mathcal{D}^{*(0,i)}$
are greatly simplified as
\begin{equation}
\mathcal{C}^{*k(0,i)}=0,\quad\mathcal{D}^{*(0,i)}=0.\label{degcon2_b2_c2_calCD}
\end{equation}

Recalling that $\dot{\phi}_{1}=\mathcal{A}\partial_{i}\varPi^{i}+\mathcal{B}(\varPi^{i})^{2}+\mathcal{C}_{i}^{(0,i)}\varPi^{i}+\mathcal{D}^{(0,i)}$
and $[\dot{\phi}_{1}(\vec{x}),H_{\mathrm{C}}]=\mathcal{C}_{k}^{*(0,i)}\varPi^{k}+\mathcal{D}^{*(0,i)}$,
(\ref{degcon2_b2_c2_calCD}) implies that $\ddot{\phi}_{1}=\mathcal{C}_{k}^{*(0,i)}\varPi^{k}+\mathcal{D}^{*(0,i)}$
vanishes identically. Consequently, no additional constraints arise,
and the Dirac matrix takes the form
\begin{equation}
\bar{A}^{ab}(\vec{x},\vec{y})=\begin{array}{|c|cc|}
\hline [\cdot,\cdot] & \phi_{1} & \dot{\phi}_{1}\\
\hline \phi_{1} & 0 & 0\\
\dot{\phi}_{1} & 0 & X
\\\hline \end{array}.
\end{equation}
On the other hand, if $[\dot{\phi}_{1}(\vec{x}),\dot{\phi}_{1}(\vec{y})]$
given in (\ref{PB_dfdf}) does not vanish weakly, the number of DOFs
is still 2.5 instead of 2. Therefore, we must further require that
$[\dot{\phi}_{1}(\vec{x}),\dot{\phi}_{1}(\vec{y})]\approx0$, which
implies that
\[
\frac{\partial f_{m,n}}{\partial A_{0}}=0.
\]
Together with the assumption (\ref{b2c2_asp}), this result indicates
that the coefficients $f_{m,n}$ must be constants.

Finally, the Dirac matrix becomes
\begin{equation}
\tilde{A}^{ab}(\vec{x},\vec{y})=\begin{array}{|c|cc|}
\hline [\cdot,\cdot] & \phi_{1} & \dot{\phi}_{1}\\
\hline \phi_{1} & 0 & 0\\
\dot{\phi}_{1} & 0 & 0
\\\hline \end{array},
\end{equation}
which indicates that the theory possesses two first-class constraints,
with the number of DOFs given by
\begin{equation}
\#_{\mathrm{DOF}}=\frac{2\times4_{\mathrm{var}}-2\times2_{1\mathrm{st}}}{2}=2.
\end{equation}
We refer to theories in this case as ``type-III'' theories. The
constraint structure in this case coincides exactly with that of the
standard Maxwell theory. We refer to Appendix \ref{app:HamMxw} for
a brief reminder of the Hamiltonian formalism for Maxwell theory.
However, due to the breaking of Lorentz invariance, the Lagrangians
can be different from that of the Maxwell theory.

At this point, it is interesting to check the gauge transformation
generated by the two first-class constraints $\phi_{1}$ and $\dot{\phi}_{1}$.
Since all coefficients are constants in this case, $\dot{\phi}_{1}$
reduces to a Gauss-type constraint $\dot{\phi}_{1}=\mathcal{A}\partial_{i}\varPi^{i}\approx$
with constant $\mathcal{A}$. A convenient choice of the generator
of gauge transformation is
\[
G[\epsilon]=\int\mathrm{d}^{3}x\left(-\dot{\epsilon}(\vec{x})\phi_{1}(\vec{x})+\frac{1}{\mathcal{A}}\epsilon(\vec{x})\dot{\phi}_{1}(\vec{x})\right),
\]
where $\epsilon(\vec{x})$ is the infinitesimal parameter. The transformations
are thus given by
\begin{align*}
\delta A_{0}(\vec{x}) & =\left[A_{0}(\vec{x}),G[\epsilon]\right]=-\dot{\epsilon}(\vec{x}),\\
\delta A_{i}(\vec{x}) & =\left[A_{i}(\vec{x}),G[\epsilon]\right]=\partial_{i}\epsilon(\vec{x}).
\end{align*}
This makes explicit the close analogy to electromagnetism, where the
Gauss-type constraint generates $\delta A_{i}=\partial_{i}\epsilon$.

After some manipulations, we can determine the Lagrangians satisfying
the second degeneracy condition. For $\mathcal{L}_{2}=f_{2}$, we
find it must be a constant, namely,
\begin{equation}
\mathcal{L}_{2}=c_{2},\label{Lag2_b2c2}
\end{equation}
which is thus trivial. For $\mathcal{L}_{3}$, both $f_{3,1}$ and
$f_{3,2}$ are constants, allowing us to write
\begin{equation}
\mathcal{L}_{3}=c_{3,1}\dot{A}_{0}+c_{3,2}\partial_{i}A^{i}\simeq0,\label{Lag3_b2c2}
\end{equation}
where we denote $f_{3,1}=c_{3,1}$ and $f_{3,2}=c_{3,2}$. Clearly
$\mathcal{L}_{3}$ is also trivial since it is a sum of total derivatives.
The only nontrivial Lagrangian is $\mathcal{L}_{4}$. We find that
$h_{4,1}$, $h_{4,2}$, $h_{4,3}$ and $f_{4,2}$ must be constants
satisfying
\begin{equation}
f_{4,2}(f_{4,2}+h_{4,3})=4\left(h_{4,2}h_{4,1}-\frac{(h_{4,3})^{2}}{4}\right).
\end{equation}
We may denote
\begin{align}
h_{4,1} & =\alpha_{3},\quad f_{4,2}=\beta_{3},\quad h_{4,3}=\gamma_{3},\\
h_{4,2} & =\frac{(\beta_{3}+\gamma_{3})^{2}-\beta_{3}\gamma_{3}}{4\alpha_{3}}.
\end{align}
We also find that $f_{4,3}$, $f_{4,4}$, and $f_{4,5}$ are constants.
Finally, we obtain
\begin{align}
\mathcal{L}_{4}\simeq & \left(\beta_{3}+\gamma_{3}\right)\dot{A}_{i}\partial^{i}A_{0}+\alpha_{3}\dot{A}_{i}\dot{A}^{i}+\frac{(\beta_{3}+\gamma_{3})^{2}-\beta_{3}\gamma_{3}}{4\alpha_{3}}\partial_{i}A_{0}\partial^{i}A_{0}\nonumber \\
 & +\left(c_{4,3}+c_{4,5}\right)\partial_{i}A_{j}\partial^{j}A^{i}+c_{4,4}\partial_{i}A_{j}\partial^{i}A^{j},\label{Lag4_b2c2}
\end{align}
where we denote $f_{4,3}=c_{4,3}$, $f_{4,4}=c_{4,4}$ and $f_{4,5}=c_{4,5}$. 

As a special case, when the system restores Lorentz invariance, the
Lagrangian $\mathcal{L}_{4}$ reduces to
\[
\mathcal{L}_{4}\propto\dot{A}_{i}\dot{A}^{i}-\partial_{i}A_{j}\partial^{i}A^{j}+\partial_{i}A_{j}\partial^{j}A^{i}+\partial_{i}A_{0}\partial^{i}A_{0}-2\dot{A}_{i}\partial^{i}A_{0},
\]
which is nothing but the Maxwell theory.

\section{Conclusion \label{sec:con}}

Recently, there has been growing interest in applying vector fields
to cosmological and gravitational-wave model building. In general,
breaking Lorentz symmetry and $\mathrm{U}(1)$ gauge invariance introduces
an additional longitudinal mode, in addition to the two usual transverse
modes of the vector field. In this work, we explored a class of spatially
covariant vector field theories and investigated the conditions required
to eliminate this extra degree of freedom so that only two modes propagate.

In Sec. \ref{sec:vfsc}, we introduced the spatially covariant vector
field theory. As a first attempt to construct a vector field theory
with two DOFs, we turned off the effects of gravity and focused on
a flat background. We assumed that the Lagrangians are polynomials
built from the first derivatives of the vector field. We then performed
a Hamiltonian analysis in Sec. \ref{sec:hampri}, demonstrating that
the theory generally propagates three DOFs unless specific conditions
are imposed on the structure of the Lagrangian. 

We found that the complete removal of the unwanted mode requires two
degeneracy conditions. The first condition is derived in Sec. \ref{sec:dgncon1}
and given in (\ref{firstdgc}), which identifies the Lagrangians in
(\ref{Lag2_dg1}), (\ref{Lag3_dg1}) and (\ref{Lag4_dg1}). If only
the first degeneracy condition is satisfied, the kernel $B_{ab}(\vec{x},\vec{y})$
defined in (\ref{Bab}) is invertible, which means all three constraints
$\{\phi_{1},\dot{\phi}_{1},\ddot{\phi}_{1}\}$ are second-class. As
a result, the theory typically propagates 2.5 DOFs which is allowed
in Lorentz-violating field theories. To fully eliminate the residual
half DOF, the second degeneracy condition is necessary. This is given
by (\ref{detBab_dg}), where the Dirac matrix $B_{ab}$ is defined
in (\ref{Bab}). There are two branches of solutions to (\ref{detBab_dg}),
denoted as the first and second branches, expressed respectively in
(\ref{degcon2_1}) and (\ref{degcon2_2}). 

In Sec. \ref{sec:dgncon2}, we analyzed both branches in detail, identifying
three distinct types of consistent theories. For the first branch,
we found two cases of solutions in Sec. \ref{subsec:branch1}. The
corresponding Lagrangians are given in (\ref{Lag2_b1c1}), (\ref{Lag3_b1c1})
and (\ref{Lag4_b1c1}) for the first case, and in (\ref{Lag2_b1c2}),
(\ref{Lag3_b1c2}) and (\ref{Lag4_b1c2}) for the second case. In
both cases, the theories possess one first-class and two second-class
constraints, which we refer to as ``type-I'' theories. For the second
branch discussed in Sec. \ref{subsec:branch2}, we identified two
special cases and thus two types of theories. The ``type-II'' theory
has four second-class constraints, with nontrivial Lagrangians given
in (\ref{Lag3_b2c1}) and (\ref{Lag4_b2c1}). The ``type-III'' theory
admits two first-class constraints, and the only nontrivial Lagrangian
is $\mathcal{L}_{4}$ given in (\ref{Lag4_b2c2}). Interestingly,
the type-III theory encompasses the Maxwell theory as a special limit
when Lorentz invariance is restored.

Our results show the possibility of constructing spatially covariant
vector field theories that propagate only two degrees of freedom,
providing a foundation for future developments in Lorentz-violating
field theories and their potential cosmological applications. In this
work, we considered only first-order derivatives of the vector field
and focused on a flat background. It will be interesting to extend
the analysis by including higher-order derivatives of the vector field
and incorporating the effects of gravity. Moreover, a systematic extension
to curved backgrounds can be carried out by promoting the building
blocks to a general foliated spacetime and replacing partial derivatives
by spatially covariant derivatives. We will explore these issues in
future work.
\begin{acknowledgments}
X.G. is supported by the National Natural Science Foundation of China
(NSFC) under Grants No. 12475068 and No. 11975020 and the Guangdong
Basic and Applied Basic Research Foundation under Grant No. 2025A1515012977.
\end{acknowledgments}

\appendix

\section{Hamiltonian formalism for Maxwell theory \label{app:HamMxw}}

Let us review the Hamiltonian formalism for Maxwell theory
\[
\mathcal{L}=-\frac{1}{4}F_{\mu\nu}F^{\mu\nu},
\]
where $F_{\mu\nu}=2\partial_{[\mu}A_{\nu]}$. 

The momenta conjugate to $A^{i}$ are $\varPi^{i}=\dot{A}^{i}-\partial^{i}A_{0}$.
The absence of a kinetic term for the temporal component $\dot{A}_{0}$
leads to the primary constraint
\[
C_{1}\equiv\varPi^{0}\approx0.
\]
The total Hamiltonian is
\[
H_{\mathrm{T}}=\int\mathrm{d}^{3}x(\mathcal{H}_{\mathrm{C}}+\lambda C_{1}),
\]
where 
\[
\mathcal{H}_{\mathrm{C}}=\frac{1}{2}\varPi^{i}\varPi_{i}+\frac{1}{2}F_{ij}F^{ij}-A_{0}\partial_{i}\varPi^{i},
\]
and $\lambda$ is an undetermined Lagrange multiplier.

The consistency condition for $\dot{C}_{1}\approx0$ gives
\[
\dot{C}_{1}(\vec{x})=[\varPi^{0}(\vec{x}),H_{C}]=\partial_{i}\varPi^{i}\equiv C_{2}(\vec{x})\approx0,
\]
which yields the secondary constraint $C_{2}\approx0$. The consistency
condition for $C_{2}\approx0$ is automatically satisfied, since
\[
\dot{C}_{2}(\vec{x})=\left[\frac{\partial\varPi^{i}(\vec{x})}{\partial x^{i}},H_{\mathrm{C}}\right]=0.
\]
Therefore $C_{1}\approx0$ and $C_{2}\approx0$ are all the constraints
in Maxwell theory.

It is straightforward to calculate the Poisson brackets among these
constraints, yielding the Dirac matrix for Maxwell theory:
\[
\begin{array}{|c|cc|}
\hline [\cdot,\cdot] & C_{1} & C_{2}\\
\hline C_{1} & 0 & 0\\
C_{2} & 0 & 0
\\\hline \end{array}.
\]
Thus, both constraints are first-class. The number of DOFs is thus
given by
\begin{equation}
\#_{\mathrm{DOF}}=\frac{2\times4_{\mathrm{var}}-2\times2_{1\mathrm{st}}}{2}=2,
\end{equation}
which corresponds to the two transverse polarization modes of the
photon.

\providecommand{\href}[2]{#2}\begingroup\raggedright\endgroup


\begin{thebibliography}{10}
	
	\bibitem{Lovelock:1971yv}
	D.~Lovelock, \emph{{The Einstein tensor and its generalizations}},
	\href{https://doi.org/10.1063/1.1665613}{\emph{J.Math.Phys.} {\bfseries 12}
		(1971) 498}.
	
	\bibitem{Horndeski:1974wa}
	G.W.~Horndeski, \emph{{Second-order scalar-tensor field equations in a
			four-dimensional space}},
	\href{https://doi.org/10.1007/BF01807638}{\emph{Int.J.Theor.Phys.} {\bfseries
			10} (1974) 363}.
	
	\bibitem{Nicolis:2008in}
	A.~Nicolis, R.~Rattazzi and E.~Trincherini, \emph{{The Galileon as a local
			modification of gravity}},
	\href{https://doi.org/10.1103/PhysRevD.79.064036}{\emph{Phys.Rev.} {\bfseries
			D79} (2009) 064036} [\href{https://arxiv.org/abs/0811.2197}{{\ttfamily
			0811.2197}}].
	
	\bibitem{Deffayet:2011gz}
	C.~Deffayet, X.~Gao, D.~Steer and G.~Zahariade, \emph{{From k-essence to
			generalised Galileons}},
	\href{https://doi.org/10.1103/PhysRevD.84.064039}{\emph{Phys.Rev.} {\bfseries
			D84} (2011) 064039} [\href{https://arxiv.org/abs/1103.3260}{{\ttfamily
			1103.3260}}].
	
	\bibitem{Kobayashi:2011nu}
	T.~Kobayashi, M.~Yamaguchi and J.~Yokoyama, \emph{{Generalized G-inflation:
			Inflation with the most general second-order field equations}},
	\href{https://doi.org/10.1143/PTP.126.511}{\emph{Prog.Theor.Phys.} {\bfseries
			126} (2011) 511} [\href{https://arxiv.org/abs/1105.5723}{{\ttfamily
			1105.5723}}].
	
	\bibitem{Langlois:2015cwa}
	D.~Langlois and K.~Noui, \emph{{Degenerate higher derivative theories beyond
			Horndeski: evading the Ostrogradski instability}},
	\href{https://doi.org/10.1088/1475-7516/2016/02/034}{\emph{JCAP} {\bfseries
			1602} (2016) 034} [\href{https://arxiv.org/abs/1510.06930}{{\ttfamily
			1510.06930}}].
	
	\bibitem{LIGOScientific:2017vwq}
	{\scshape LIGO Scientific, Virgo} collaboration, \emph{{GW170817: Observation
			of Gravitational Waves from a Binary Neutron Star Inspiral}},
	\href{https://doi.org/10.1103/PhysRevLett.119.161101}{\emph{Phys. Rev. Lett.}
		{\bfseries 119} (2017) 161101}
	[\href{https://arxiv.org/abs/1710.05832}{{\ttfamily 1710.05832}}].
	
	\bibitem{Bettoni:2016mij}
	D.~Bettoni, J.M.~Ezquiaga, K.~Hinterbichler and M.~Zumalacárregui,
	\emph{{Speed of Gravitational Waves and the Fate of Scalar-Tensor Gravity}},
	\href{https://doi.org/10.1103/PhysRevD.95.084029}{\emph{Phys. Rev.}
		{\bfseries D95} (2017) 084029}
	[\href{https://arxiv.org/abs/1608.01982}{{\ttfamily 1608.01982}}].
	
	\bibitem{Ezquiaga:2017ekz}
	J.M.~Ezquiaga and M.~Zumalacárregui, \emph{{Dark Energy After GW170817: Dead
			Ends and the Road Ahead}},
	\href{https://doi.org/10.1103/PhysRevLett.119.251304}{\emph{Phys. Rev. Lett.}
		{\bfseries 119} (2017) 251304}
	[\href{https://arxiv.org/abs/1710.05901}{{\ttfamily 1710.05901}}].
	
	\bibitem{Sakstein:2017xjx}
	J.~Sakstein and B.~Jain, \emph{{Implications of the Neutron Star Merger
			GW170817 for Cosmological Scalar-Tensor Theories}},
	\href{https://doi.org/10.1103/PhysRevLett.119.251303}{\emph{Phys. Rev. Lett.}
		{\bfseries 119} (2017) 251303}
	[\href{https://arxiv.org/abs/1710.05893}{{\ttfamily 1710.05893}}].
	
	\bibitem{Baker:2017hug}
	T.~Baker, E.~Bellini, P.G.~Ferreira, M.~Lagos, J.~Noller and I.~Sawicki,
	\emph{{Strong constraints on cosmological gravity from GW170817 and GRB
			170817A}}, \href{https://doi.org/10.1103/PhysRevLett.119.251301}{\emph{Phys.
			Rev. Lett.} {\bfseries 119} (2017) 251301}
	[\href{https://arxiv.org/abs/1710.06394}{{\ttfamily 1710.06394}}].
	
	\bibitem{Creminelli:2017sry}
	P.~Creminelli and F.~Vernizzi, \emph{{Dark Energy after GW170817 and
			GRB170817A}},
	\href{https://doi.org/10.1103/PhysRevLett.119.251302}{\emph{Phys. Rev. Lett.}
		{\bfseries 119} (2017) 251302}
	[\href{https://arxiv.org/abs/1710.05877}{{\ttfamily 1710.05877}}].
	
	\bibitem{Tasinato:2014eka}
	G.~Tasinato, \emph{{Cosmic Acceleration from Abelian Symmetry Breaking}},
	\href{https://doi.org/10.1007/JHEP04(2014)067}{\emph{JHEP} {\bfseries 04}
		(2014) 067} [\href{https://arxiv.org/abs/1402.6450}{{\ttfamily 1402.6450}}].
	
	\bibitem{Heisenberg:2014rta}
	L.~Heisenberg, \emph{{Generalization of the Proca Action}},
	\href{https://doi.org/10.1088/1475-7516/2014/05/015}{\emph{JCAP} {\bfseries
			1405} (2014) 015} [\href{https://arxiv.org/abs/1402.7026}{{\ttfamily
			1402.7026}}].
	
	\bibitem{Allys:2015sht}
	E.~Allys, P.~Peter and Y.~Rodriguez, \emph{{Generalized Proca action for an
			Abelian vector field}},
	\href{https://doi.org/10.1088/1475-7516/2016/02/004}{\emph{JCAP} {\bfseries
			02} (2016) 004} [\href{https://arxiv.org/abs/1511.03101}{{\ttfamily
			1511.03101}}].
	
	\bibitem{DeFelice:2016yws}
	A.~De~Felice, L.~Heisenberg, R.~Kase, S.~Mukohyama, S.~Tsujikawa and
	Y.-l.~Zhang, \emph{{Cosmology in generalized Proca theories}},
	\href{https://doi.org/10.1088/1475-7516/2016/06/048}{\emph{JCAP} {\bfseries
			06} (2016) 048} [\href{https://arxiv.org/abs/1603.05806}{{\ttfamily
			1603.05806}}].
	
	\bibitem{Heisenberg:2016eld}
	L.~Heisenberg, R.~Kase and S.~Tsujikawa, \emph{{Beyond generalized Proca
			theories}}, \href{https://doi.org/10.1016/j.physletb.2016.07.052}{\emph{Phys.
			Lett. B} {\bfseries 760} (2016) 617}
	[\href{https://arxiv.org/abs/1605.05565}{{\ttfamily 1605.05565}}].
	
	\bibitem{DeFelice:2016uil}
	A.~De~Felice, L.~Heisenberg, R.~Kase, S.~Mukohyama, S.~Tsujikawa and
	Y.-l.~Zhang, \emph{{Effective gravitational couplings for cosmological
			perturbations in generalized Proca theories}},
	\href{https://doi.org/10.1103/PhysRevD.94.044024}{\emph{Phys. Rev. D}
		{\bfseries 94} (2016) 044024}
	[\href{https://arxiv.org/abs/1605.05066}{{\ttfamily 1605.05066}}].
	
	\bibitem{BeltranJimenez:2016rff}
	J.~Beltran~Jimenez and L.~Heisenberg, \emph{{Derivative self-interactions for a
			massive vector field}},
	\href{https://doi.org/10.1016/j.physletb.2016.04.017}{\emph{Phys. Lett. B}
		{\bfseries 757} (2016) 405}
	[\href{https://arxiv.org/abs/1602.03410}{{\ttfamily 1602.03410}}].
	
	\bibitem{DeFelice:2016cri}
	A.~De~Felice, L.~Heisenberg, R.~Kase, S.~Tsujikawa, Y.-l.~Zhang and G.-B.~Zhao,
	\emph{{Screening fifth forces in generalized Proca theories}},
	\href{https://doi.org/10.1103/PhysRevD.93.104016}{\emph{Phys. Rev. D}
		{\bfseries 93} (2016) 104016}
	[\href{https://arxiv.org/abs/1602.00371}{{\ttfamily 1602.00371}}].
	
	\bibitem{Chagoya:2016aar}
	J.~Chagoya, G.~Niz and G.~Tasinato, \emph{{Black Holes and Abelian Symmetry
			Breaking}},
	\href{https://doi.org/10.1088/0264-9381/33/17/175007}{\emph{Class. Quant.
			Grav.} {\bfseries 33} (2016) 175007}
	[\href{https://arxiv.org/abs/1602.08697}{{\ttfamily 1602.08697}}].
	
	\bibitem{Babichev:2017rti}
	E.~Babichev, C.~Charmousis and M.~Hassaine, \emph{{Black holes and solitons in
			an extended Proca theory}},
	\href{https://doi.org/10.1007/JHEP05(2017)114}{\emph{JHEP} {\bfseries 05}
		(2017) 114} [\href{https://arxiv.org/abs/1703.07676}{{\ttfamily
			1703.07676}}].
	
	\bibitem{Heisenberg:2017hwb}
	L.~Heisenberg, R.~Kase, M.~Minamitsuji and S.~Tsujikawa, \emph{{Black holes in
			vector-tensor theories}},
	\href{https://doi.org/10.1088/1475-7516/2017/08/024}{\emph{JCAP} {\bfseries
			08} (2017) 024} [\href{https://arxiv.org/abs/1706.05115}{{\ttfamily
			1706.05115}}].
	
	\bibitem{Heisenberg:2017xda}
	L.~Heisenberg, R.~Kase, M.~Minamitsuji and S.~Tsujikawa, \emph{{Hairy
			black-hole solutions in generalized Proca theories}},
	\href{https://doi.org/10.1103/PhysRevD.96.084049}{\emph{Phys. Rev. D}
		{\bfseries 96} (2017) 084049}
	[\href{https://arxiv.org/abs/1705.09662}{{\ttfamily 1705.09662}}].
	
	\bibitem{Minamitsuji:2021gcq}
	M.~Minamitsuji, \emph{{Black holes in the quadratic-order extended
			vector-tensor theories}},
	\href{https://doi.org/10.1088/1361-6382/abed62}{\emph{Class. Quant. Grav.}
		{\bfseries 38} (2021) 105011}
	[\href{https://arxiv.org/abs/2105.08936}{{\ttfamily 2105.08936}}].
	
	\bibitem{Garcia-Saenz:2021uyv}
	S.~Garcia-Saenz, A.~Held and J.~Zhang, \emph{{Destabilization of Black Holes
			and Stars by Generalized Proca Fields}},
	\href{https://doi.org/10.1103/PhysRevLett.127.131104}{\emph{Phys. Rev. Lett.}
		{\bfseries 127} (2021) 131104}
	[\href{https://arxiv.org/abs/2104.08049}{{\ttfamily 2104.08049}}].
	
	\bibitem{Dong:2023xyb}
	Y.-Q.~Dong, Y.-Q.~Liu and Y.-X.~Liu, \emph{{Polarization modes of gravitational
			waves in generalized Proca theory}},
	\href{https://doi.org/10.1103/PhysRevD.109.024014}{\emph{Phys. Rev. D}
		{\bfseries 109} (2024) 024014}
	[\href{https://arxiv.org/abs/2305.12516}{{\ttfamily 2305.12516}}].
	
	\bibitem{Lai:2024fza}
	X.-B.~Lai, Y.-Q.~Dong, Y.-Q.~Liu and Y.-X.~Liu, \emph{{Polarization modes of
			gravitational waves in general Einstein-vector theory}},
	\href{https://doi.org/10.1103/PhysRevD.110.064073}{\emph{Phys. Rev. D}
		{\bfseries 110} (2024) 064073}
	[\href{https://arxiv.org/abs/2405.20577}{{\ttfamily 2405.20577}}].
	
	\bibitem{Dong:2024zal}
	Y.-Q.~Dong, X.-B.~Lai, Y.-Q.~Liu and Y.-X.~Liu, \emph{{Gravitational-wave
			effects in the most general vector{\textendash}tensor theory}},
	\href{https://doi.org/10.1140/epjc/s10052-025-14378-5}{\emph{Eur. Phys. J. C}
		{\bfseries 85} (2025) 645}
	[\href{https://arxiv.org/abs/2409.11838}{{\ttfamily 2409.11838}}].
	
	\bibitem{Heisenberg:2018vsk}
	L.~Heisenberg, \emph{{A systematic approach to generalisations of General
			Relativity and their cosmological implications}},
	\href{https://doi.org/10.1016/j.physrep.2018.11.006}{\emph{Phys. Rept.}
		{\bfseries 796} (2019) 1} [\href{https://arxiv.org/abs/1807.01725}{{\ttfamily
			1807.01725}}].
	
	\bibitem{Kimura:2016rzw}
	R.~Kimura, A.~Naruko and D.~Yoshida, \emph{{Extended vector-tensor theories}},
	\href{https://doi.org/10.1088/1475-7516/2017/01/002}{\emph{JCAP} {\bfseries
			1701} (2017) 002} [\href{https://arxiv.org/abs/1608.07066}{{\ttfamily
			1608.07066}}].
	
	\bibitem{GallegoCadavid:2019zke}
	A.~Gallego~Cadavid and Y.~Rodriguez, \emph{{A systematic procedure to build the
			beyond generalized Proca field theory}},
	\href{https://doi.org/10.1016/j.physletb.2019.134958}{\emph{Phys. Lett. B}
		{\bfseries 798} (2019) 134958}
	[\href{https://arxiv.org/abs/1905.10664}{{\ttfamily 1905.10664}}].
	
	\bibitem{GallegoCadavid:2020dho}
	A.~Gallego~Cadavid, Y.~Rodriguez and L.G.~G\'omez, \emph{{Generalized SU(2)
			Proca theory reconstructed and beyond}},
	\href{https://doi.org/10.1103/PhysRevD.102.104066}{\emph{Phys. Rev. D}
		{\bfseries 102} (2020) 104066}
	[\href{https://arxiv.org/abs/2009.03241}{{\ttfamily 2009.03241}}].
	
	\bibitem{GallegoCadavid:2021ljh}
	A.~Gallego~Cadavid, C.M.~Nieto and Y.~Rodriguez, \emph{{Towards the extended
			SU(2) Proca theory}},
	\href{https://doi.org/10.1103/PhysRevD.105.124060}{\emph{Phys. Rev. D}
		{\bfseries 105} (2022) 124060}
	[\href{https://arxiv.org/abs/2110.14623}{{\ttfamily 2110.14623}}].
	
	\bibitem{deRham:2021efp}
	C.~de~Rham, S.~Garcia-Saenz, L.~Heisenberg and V.~Pozsgay, \emph{{Cosmology of
			Extended Proca-Nuevo}},
	\href{https://doi.org/10.1088/1475-7516/2022/03/053}{\emph{JCAP} {\bfseries
			03} (2022) 053} [\href{https://arxiv.org/abs/2110.14327}{{\ttfamily
			2110.14327}}].
	
	\bibitem{Aoki:2021wew}
	K.~Aoki, M.A.~Gorji, S.~Mukohyama and K.~Takahashi, \emph{{The effective field
			theory of vector-tensor theories}},
	\href{https://doi.org/10.1088/1475-7516/2022/01/059}{\emph{JCAP} {\bfseries
			01} (2022) 059} [\href{https://arxiv.org/abs/2111.08119}{{\ttfamily
			2111.08119}}].
	
	\bibitem{Aoki:2023bmz}
	K.~Aoki, M.A.~Gorji, S.~Mukohyama, K.~Takahashi and V.~Yingcharoenrat,
	\emph{{Effective field theory of black hole perturbations in vector-tensor
			gravity}}, \href{https://doi.org/10.1088/1475-7516/2024/03/012}{\emph{JCAP}
		{\bfseries 03} (2024) 012}
	[\href{https://arxiv.org/abs/2311.06767}{{\ttfamily 2311.06767}}].
	
	\bibitem{Deffayet:2013tca}
	C.~Deffayet, A.E.~Gumrukcuoglu, S.~Mukohyama and Y.~Wang, \emph{{A no-go
			theorem for generalized vector Galileons on flat spacetime}},
	\href{https://doi.org/10.1007/JHEP04(2014)082}{\emph{JHEP} {\bfseries 1404}
		(2014) 082} [\href{https://arxiv.org/abs/1312.6690}{{\ttfamily 1312.6690}}].
	
	\bibitem{Colleaux:2023cqu}
	A.~Coll\'eaux, D.~Langlois and K.~Noui, \emph{{Classification of generalised
			higher-order Einstein-Maxwell Lagrangians}},
	\href{https://doi.org/10.1007/JHEP03(2024)041}{\emph{JHEP} {\bfseries 03}
		(2024) 041} [\href{https://arxiv.org/abs/2312.14814}{{\ttfamily
			2312.14814}}].
	
	\bibitem{Colleaux:2024ndy}
	A.~Coll\'eaux, D.~Langlois and K.~Noui, \emph{{Degenerate higher-order Maxwell
			theories in flat space-time}},
	\href{https://doi.org/10.1007/JHEP10(2024)218}{\emph{JHEP} {\bfseries 10}
		(2024) 218} [\href{https://arxiv.org/abs/2404.18715}{{\ttfamily
			2404.18715}}].
	
	\bibitem{Colleaux:2025vtm}
	A.~Coll{\'e}aux and K.~Noui, \emph{{Degenerate higher-order Maxwell-Einstein
			theories}}, \href{https://doi.org/10.1007/JHEP10(2025)007}{\emph{JHEP}
		{\bfseries 10} (2025) 007}
	[\href{https://arxiv.org/abs/2502.03311}{{\ttfamily 2502.03311}}].
	
	\bibitem{Afshordi:2006ad}
	N.~Afshordi, D.J.H.~Chung and G.~Geshnizjani, \emph{{Cuscuton: A Causal Field
			Theory with an Infinite Speed of Sound}},
	\href{https://doi.org/10.1103/PhysRevD.75.083513}{\emph{Phys. Rev.}
		{\bfseries D75} (2007) 083513}
	[\href{https://arxiv.org/abs/hep-th/0609150}{{\ttfamily hep-th/0609150}}].
	
	\bibitem{Lin:2017oow}
	C.~Lin and S.~Mukohyama, \emph{{A Class of Minimally Modified Gravity
			Theories}}, \href{https://doi.org/10.1088/1475-7516/2017/10/033}{\emph{JCAP}
		{\bfseries 1710} (2017) 033}
	[\href{https://arxiv.org/abs/1708.03757}{{\ttfamily 1708.03757}}].
	
	\bibitem{Gao:2019twq}
	X.~Gao and Z.-B.~Yao, \emph{{Spatially covariant gravity theories with two
			tensorial degrees of freedom: the formalism}},
	\href{https://doi.org/10.1103/PhysRevD.101.064018}{\emph{Phys. Rev.}
		{\bfseries D101} (2020) 064018}
	[\href{https://arxiv.org/abs/1910.13995}{{\ttfamily 1910.13995}}].
	
	\bibitem{Sanongkhun:2019ntn}
	J.~Sanongkhun and P.~Vanichchapongjaroen, \emph{{On constrained analysis and
			diffeomorphism invariance of generalised Proca theories}},
	\href{https://doi.org/10.1007/s10714-020-02678-y}{\emph{Gen. Rel. Grav.}
		{\bfseries 52} (2020) 26} [\href{https://arxiv.org/abs/1907.12794}{{\ttfamily
			1907.12794}}].
	
	\bibitem{ErrastiDiez:2019ttn}
	V.~Errasti~D{\'\i}ez, B.~Gording, J.A.~M{\'e}ndez-Zavaleta and A.~Schmidt-May,
	\emph{{Complete theory of Maxwell and Proca fields}},
	\href{https://doi.org/10.1103/PhysRevD.101.045008}{\emph{Phys. Rev. D}
		{\bfseries 101} (2020) 045008}
	[\href{https://arxiv.org/abs/1905.06967}{{\ttfamily 1905.06967}}].
	
	\bibitem{ErrastiDiez:2019trb}
	V.~Errasti~D{\'\i}ez, B.~Gording, J.A.~M{\'e}ndez-Zavaleta and A.~Schmidt-May,
	\emph{{Maxwell-Proca theory: Definition and construction}},
	\href{https://doi.org/10.1103/PhysRevD.101.045009}{\emph{Phys. Rev. D}
		{\bfseries 101} (2020) 045009}
	[\href{https://arxiv.org/abs/1905.06968}{{\ttfamily 1905.06968}}].
	
	\bibitem{Floreanini:1987as}
	R.~Floreanini and R.~Jackiw, \emph{{Selfdual Fields as Charge Density
			Solitons}}, \href{https://doi.org/10.1103/PhysRevLett.59.1873}{\emph{Phys.
			Rev. Lett.} {\bfseries 59} (1987) 1873}.
	
	\bibitem{Abreu:2004kn}
	E.M.C.~Abreu and C.~Wotzasek, \emph{{Topics on the quantum dynamics of chiral
			bosons}},  \href{https://arxiv.org/abs/hep-th/0410019}{{\ttfamily
			hep-th/0410019}}.
	
	\bibitem{Blas:2009yd}
	D.~Blas, O.~Pujolas and S.~Sibiryakov, \emph{{On the Extra Mode and
			Inconsistency of Horava Gravity}},
	\href{https://doi.org/10.1088/1126-6708/2009/10/029}{\emph{JHEP} {\bfseries
			0910} (2009) 029} [\href{https://arxiv.org/abs/0906.3046}{{\ttfamily
			0906.3046}}].
	
	\bibitem{Gao:2018znj}
	X.~Gao and Z.-B.~Yao, \emph{{Spatially covariant gravity with velocity of the
			lapse function: the Hamiltonian analysis}},
	\href{https://doi.org/10.1088/1475-7516/2019/05/024}{\emph{JCAP} {\bfseries
			1905} (2019) 024} [\href{https://arxiv.org/abs/1806.02811}{{\ttfamily
			1806.02811}}].
	
	\bibitem{Yao:2020tur}
	Z.-B.~Yao, M.~Oliosi, X.~Gao and S.~Mukohyama, \emph{{Minimally modified
			gravity with an auxiliary constraint: A Hamiltonian construction}},
	\href{https://doi.org/10.1103/PhysRevD.103.024032}{\emph{Phys. Rev. D}
		{\bfseries 103} (2021) 024032}
	[\href{https://arxiv.org/abs/2011.00805}{{\ttfamily 2011.00805}}].
	
	\bibitem{Iyonaga:2020bmm}
	A.~Iyonaga, K.~Takahashi and T.~Kobayashi, \emph{{Extended Cuscuton as Dark
			Energy}}, \href{https://doi.org/10.1088/1475-7516/2020/07/004}{\emph{JCAP}
		{\bfseries 07} (2020) 004}
	[\href{https://arxiv.org/abs/2003.01934}{{\ttfamily 2003.01934}}].
	
\end{thebibliography}
\end{document}